\documentclass{jaa}
\usepackage{natbib}
\usepackage{graphicx}
\usepackage{caption}
\usepackage{array}
\usepackage{color}

\usepackage{xltabular}
\usepackage{booktabs}
\usepackage{amsmath}
\usepackage{pdflscape}
\usepackage{multirow}

\newcolumntype{C}[1]{>{\centering\arraybackslash}m{#1}}

\newcolumntype{F}{>{\centering\arraybackslash}X}

\makeatletter
\newcommand\footnoteref[1]{\protected@xdef\@thefnmark{\ref{#1}}\@footnotemark}
\makeatother

\begin{document}\sloppy

\title{HEL1OS on \textit{Aditya-L1} Mission: Operations, Data Processing and\\ Monitoring of Sun in Hard X-rays}

\author{Ravishankar B. T.\textsuperscript{1,*},
Manju Sudhakar\textsuperscript{1},
Srikar Paavan Tadepalli\textsuperscript{1},
Kiran Kumar Kalisetti\textsuperscript{1,2},
Madhavan R. Menon\textsuperscript{3},
Anand Jain\textsuperscript{4},
Reenu Palawat\textsuperscript{1},
Brajpal Singh\textsuperscript{1},
Rethika T.\textsuperscript{4},
Priyanka Upadhyay\textsuperscript{4},
Prapti Mittal\textsuperscript{5},
Amit Purohit\textsuperscript{5},
Srikanth M.\textsuperscript{4},
and
Anuj Nandi\textsuperscript{1}}
\affilOne{\textsuperscript{1}ISRO Satellite Integration and Test Establishment (ISITE), U.R. Rao Satellite Centre (URSC), Indian Space Research Organisation (ISRO), Bengaluru - 560037, India.\\}
\affilTwo{\textsuperscript{2}Department of Physics, Indian Institute of Science, Bengaluru 560012, India.\\}
\affilThree{\textsuperscript{3}Shiv Nadar Institute of Eminence, Greater Noida - 201314, India\\}
\affilFour{\textsuperscript{4}U.R. Rao Satellite Centre (URSC), Indian Space Research Organisation (ISRO), Bengaluru - 560017, India.\\}
\affilFive{\textsuperscript{5}ISRO Telemetry, Tracking and Command Network (ISTRAC), Indian Space Research Organisation (ISRO), Bengaluru - 560058, India\\}


\twocolumn[{

\maketitle

\corres{ravibt@ursc.gov.in}

\msinfo{27 July 2026}{}

\begin{abstract}
HEL1OS (High Energy L1 Orbiting X-ray Spectrometer) is the hard
X-ray spectrometer on-board \textit{Aditya-L1} solar mission,
monitoring the Sun continuously in the hard X-ray band of $8$~keV
to $150$~keV from the vantage observation platform around the
Sun-Earth Lagragian Point 1 (L1).  HEL1OS was commissioned in
end of October, 2023 and the final instrument configuration was set
by end of June, 2024 after all the performance verifications. The
instrument always operates in the event mode, and the down-linked
data is processed in an automated pipeline to generate Science-ready
products comprising mainly of solar spectra in $8$~keV to $150$~keV
energy band. The data products of observations since 
December, 2023 are hosted on the internet with open
access. This includes data from the performance
verification phase of the payload operations. In this paper,
the details of the commissioning and performance verification phases,
all the stages of the fully automated data pipeline, and the data
products of HEL1OS are discussed. Also a few scientific results are
discussed to illustrate the timing and spectral capabilities of the
instrument in monitoring Sun in the X-ray energy bands.
\end{abstract}

\keywords{fully automated data pipeline---Solar X-ray spectrometer
---Hard X-ray spectrometer.}

}]


\doinum{12.3456/s78910-011-012-3}
\artcitid{\#\#\#\#}
\volnum{000}
\year{0000}
\pgrange{1--}
\setcounter{page}{1}
\lp{1}

\section{Introduction}

The HEL1OS payload (\cite{Anuj:solphy}) on \textit{Aditya-L1}
(\cite{2017CSci..113..610S}), the first dedicated solar mission by
ISRO, is meant to monitor Sun in the hard X-ray energy band of 
$8$~keV to $150$~keV. It is mounted on the intermediate deck of the
spacecraft viewing along the +Yaw axis. The wide energy band of
$8$~keV to $150$~keV is achieved by employing two different kinds
of detectors -- a pair of Cadmium Telluride (CdTe) single pixel
detectors operating over $8$~keV -- $70$~keV, and a pair of Cadmium
Zinc Telluride (CZT) $256$~pixel detectors operating over $20$~keV --
$150$~keV. The total geometric areas of the CdTe and CZT detector
pairs are $0.5$~cm${^2}$ and $32$~cm${^2}$, respectively.

After the launch of the \textit{Aditya-L1} spacecraft on
September 2, 2023, HEL1OS was commissioned on October 27, 2023 while
\textit{Aditya-L1} was in the cruise phase towards the L1 point.
Post commissioning, the instrument performance parameters
were studied, and the operational configuration was fine-tuned for
regular operations by end of June, 2024. During this performance
verification phase the on-board logics tested and tuned include:
hot-pixel \& bunched-pixel logics of the CZT detectors, and, the
digital pulse processing (DPP) parameters of the CdTe detectors.
The hot-pixel logic of the CZT detectors rejects data from pixels
with background events higher than a set count threshold (more
details are covered in \cite{Srikar:submitted}).
The bunched-pixel logic of the CZT detectors tags and rejects
those events which are recorded within the set time-window of
$6$~$\mu$s, in the entire operating energy range.
This rejection of near-simultaneous detection in multiple pixels
may be either among adjacent pixels or also in pixels which are far
apart as well. The sources of these energetic particles may be Galatic
cosmic rays, solar energetic particles (SEP) or coronal mass ejection
(CME) events. The tuneable DPP parameters of the CdTe detectors
include the energy thresholds, peaking time, gain and pile-up
rejection. More details of these onboard modules for the CdTe and CZT
detectors are presented in \cite{Srikar:submitted}. The instrument
always functions in the final set configuration and operates in the
event-mode, where in it records every detected photon at the clock
resolution of $10$~ms cadence. The rawdata recorded on-board in
the instrument's FPGA is down-linked via the spacecraft's
data-handling and communication sub-systems, typically twice a day.
Subsequently, higher level products are generated in a fully
automated pipeline. The data products are made available immediately
for web-hosting at https://pradan1.issdc.gov.in/al1/ with open
access, without any lock-in period.

This paper provides an overview of the HEL1OS operational aspects
during the crucial commissioning and performance verification phases
(in section~\ref{sec:oper}). The architecture of the automated
pipeline operating between the Indian Space Science Data Centre
(ISSDC) and the Payload Operation Centre (POC) is described in
section~\ref{sec:arch}. A detailed account on the data processing
and generation of the data products including intermediate stages,
are provided in section~\ref{sec:lev1proc}. The information about the
dissemination through the Policy based data Retrieval, Analytics,
Dissemination And Notification (PRADAN) portal at ISSDC is presented
in section~\ref{sec:dissem}. In section~\ref{sec:hardsun}, we present
the long term X-ray lightcurve of Sun as seen by HEL1OS, and
illustrate the timing and spectral capabilities of the instrument
for a few sample detected flares. Also presented in the same section
are the characteristics of several solar flares observed by HEL1OS
over an year of its operation since July, $2024$.

\section{HEL1OS Operations}\label{sec:oper}
The operations of \textit{Aditya-L1} spacecraft
(\cite{2025SoPh..300..128P}) were carried out in three phases: Earth
bound phase, Cruise phase and L1 orbiting phase. The initial plan was
to switch ON the HEL1OS payload in the L1 phase.
However, even while in the Cruise phase, once the payload operating
parameters of temperature, etc, were achieved in the desired optimum
range, and the spacecraft pointing towards the Sun was established
within the required accuracy, the HEL1OS payload was commissioned on
October $27$, $2023$. Following are the operational sequences set for
the HEL1OS payload:

\textbf{ON sequence} \vspace{-2mm}
\begin{itemize}
    \item[1.] Payload ON \vspace{-2mm}
    \item[2.] High Voltage ON \vspace{-2mm}
    \item[3.] Configuration Parameters Uplink \vspace{-2mm}
    \item[4.] Data Acquisition Enable \vspace{-1mm}
\end{itemize}

\textbf{OFF sequence} \vspace{-2mm}
\begin{itemize}
    \item[1.] Data Acquisition Disable \vspace{-2mm}
    \item[2.] High Voltage OFF \vspace{-2mm}
    \item[3.] Payload OFF \vspace{-1mm}
\end{itemize}

HEL1OS configuration setting is via bulk mode of command uplink with
recommended inter-command delays. To take care of this, Macros, which
are onboard command structures, were designed apriori to take care of
the switch ON operations and the commissioning operations as per the
timeline. After the payload was commissioned, Performance Verification
phase (PV phase) was carried out with different configurations
uplinked to the payload. Currently, the payload is continuously ON
with fine tuned configurations. The observations with HEL1OS are not
driven by proposals and it is designed to always operate in the Event
mode during pointings towards Sun as well as towards
calibration sources.

\subsection{Commissioning and Performance Verification}

Initially, when HEL1OS was commissioned on October $27$, $2023$
(\cite{Anuj:solphy}; \cite{Srikar:submitted}), the energy threshold
values of the CZT detectors was set at $40$~keV, and those of the CdTe
detectors were set at $8$~keV. The hot-pixel and the bunched-pixel on
board logics of the CZT detectors were initially kept disabled. In the
following few weeks, the DPP parameters of the CdTe detectors were
carefully tuned for optimal spectral response. The low energy
threshold of the CZT detectors was brought down in steps from
$40$~keV, to $30$~keV, and ultimately to $20$~keV. Then the bunched
pixel logic of the CZT was also enabled which brought down the recorded
event-rate. There is a Temperature-read module of CZT as part of its
read-out application specific integrated circuit (ASIC) which
intermittently provides the values of the device temperture. However,
during these temperature-read intervals the CZT events are not read.
The Temperature-read module of the CZT was kept ON for some months to
study the variations of the device temperature in orbit. After it was
ascertained that the CZT device temperature stayed stable, and also
given that there is a known offset of CZT device temperature from the
detector tray temperature (which in turn is measured all the time),
the Temperature-read module of CZT was disabled by end of June,
$2024$. Therefore till then, the CZT ligthcurves have data gaps
intermittently when the Temperature-read module got triggered.

\begin{figure}[!h]
    \centering
    \includegraphics[width=\columnwidth, trim=4mm 4mm 4mm 4mm,clip]{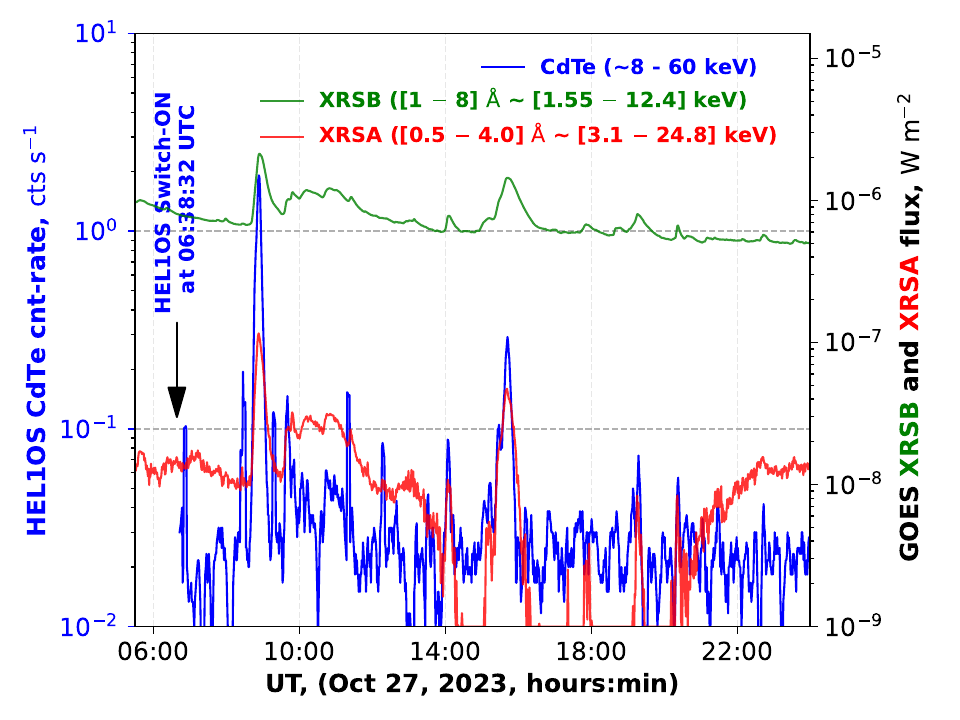}
    \caption{The lightcurve of the combined CdTe data plotted for
    October $27$, $2023$ (in blue), the day on which HEL1OS was
    switched ON in initial configuration before DPP tuning; GOES
    lightcurves of both XRSA and XRSB channels are also plotted (in
    red and green respectively) for the same time interval, for
    comparison.}
    \label{fig:day1}
    \vspace{-3mm}
\end{figure}

When the data acquistion on board the spacecraft was initiated at
$06$:$38$:$32$~UTC on October $27$, $2023$, the Geostationary
Operational Environmental Satellite (GOES) X-ray sensor (XRS) flux was
at B8 class level. Shown in Figure~\ref{fig:day1} is the lightcurve of
the combined CdTe data plotted for that first day of HEL1OS switch-on.
Also plotted are the GOES/XRS light curves in both short waveband XRSA
($3.1-24.8$~keV or $0.5-4.0$~\AA), and long waveband XRSB
($1.55-12.4$~keV or $1-8$~\AA), for comparison. The CdTe detectors have
detected the C2.1 class flare (peaking at $\sim$~$08$:$54$~UTC) as well
as C1.1 flare (peak at $\sim$~$15$:$32$~UTC), albeit with poor
statistics given that the DPP parameters of the CdTe detectors were
not yet fine-tuned then.
The $40$~keV low-energy threshold of the CZT detectors at the time
resulted in detection of insignificant number of events in them.

\subsection{Operational Normal Phase}

The operating configuration of the HEL1OS payload was set by end of
June, $2024$, and has been operating continously with these fine
tuned set of operating parameters listed in the
Table~\ref{tab:config}. The payload data recorded in the onboard
storage are played back during the spacecraft visibility of the
supporting ground station. The ISRO Deep Space Network (IDSN)
antenna at Bangalore is used for payload data dumping. The dumped
data is acquired and extracted by Payload data acquisition software
residing at the station. The acquired raw data is further time
tagged and processed with spacecraft auxiliary data information
in the Level-0 processing undertaken at ISSDC. The Level-0 data
are made available, through the data servers maintained at ISSDC
via dedicated network channels, for validation and further
processing towards data product generation at the POC (more details
are presented in section~\ref{sec:arch}).

\begin{table}[!h]
    {\small
    \caption{Operating configuration of HEL1OS as set by end of June,
    $2024$}
    \label{tab:config}
    \begin{tabular}{|C{3.2cm}|C{4.5cm}|}\hline
        \textbf{Parameter/Logic} & \textbf{Value/Status} \\\hline
        CZT Bunched-pixel logic & Enabled \\\hline
        CZT Hot-pixel rejection logic & Disabled \\\hline
        CZT Saturated events rejection logic & Enabled \\\hline
        Low Energy Threshold & CdTe detectors: $\sim8$~keV
        \hspace{5mm} CZT detectors: $\sim20$~keV \\\hline
        Operating temperature & CdTe detectors: -$40^\circ$C $-$ -$30^\circ$C
        CZT detectors: $15^\circ$C $-$ $25^\circ$C\\\hline
        CZT Bunched-pixel logic time window & $6$~$\mu$s \\\hline
        CdTe Pile-up pulse width counter & $1.6$~$\mu$s \\\hline
        Corona logic current threshold & $120$~mA \\\hline
    \end{tabular}
    \vspace{-3mm}
    } 
\end{table}

\subsubsection{Payload Safety}
The HEL1OS payload is continuously ON and continuous health monitoring
using satellite telemetry is required to handle and take actions
during the payload contingencies. In the case of \textit{Aditya-L1}
mission, continuous telemetry (TM) data is not available due to
tracking station visibility gaps. Hence, in order to safeguard the
payload during such visibility gaps, and quickly take appropriate
actions onboard in case of contingencies, onboard
safety monitoring is implemented and is always active. The safety
monitoring scheme continuously monitors the satellite health
parameters as well as payload specific critical parameters that are
available onboard, and quickly acts by initiating the Switch-OFF
sequence autonomously based on set thresholds, thereby safe-guarding
it. As part of the safety logic on board, following parameters of
HEL1OS are monitored: (a) detection of corona based on set current
threshold in each detector (mentioned in Table~\ref{tab:config})
-- if triggered, the high voltage (HV) supply for respective
detector is switched off, and (b) the temperature of any of the
detectors going beyond the set limits mentioned in
Table~\ref{tab:config} -- in this case the HV of the respective
detector is switched off. The instrument operation restoration in
case of trigger of any of these safety aspects will be by manual
intervention in uplinking appropriate telecommands after studying
all available payload parameters.

\subsection{Onboard Calibration}
HEL1OS carries two $^{241}$Am radioactive sources one for each
detector pair as detailed in \cite{Anuj:solphy}. 
The $59.54$~keV decay line from these sources is used to monitor
shifts in detector gain and resolution over time. Since commissioning,
the detector performance has remained stable; further details will be
discussed in \cite{Srikar:submitted}.
The effective area calibration will be undertaken using future planned
Crab observations. The Mission constraints allow pointing to the Crab
source only during two slots of time-intervals every year. 
\vspace{-5mm}

\section{Data Architecture}\label{sec:arch}
The HEL1OS payload raw data packets down-linked from the spacecraft
to the ground station are converted to Level-0 format after packing
the payload data with relevant auxiliary data for the observation
duration. This Level-0 data generated at ISSDC is made available at
its data server. The data polling software run by the POC transfers
the newly posted Level-0 data from the data server at ISSDC to the
POC system.
The POC system undertakes Level-0 to Level-1 processing and posts
the Level-1 data products to ISSDC for archival, storage and
dissemination. 
The science-ready Level-1 data products include the Office of Guest
Investigators Program (OGIP) specified Type-II Flexible
Image Transport System (FITS) Spectral data and the energy resolved
lightcurves. These data products are transferred to ISSDC for
dissemination on the PRADAN portal without any lock-in period.
ISSDC also ensures long term archival of the data products.
The data transfer between ISSDC data server and the POC system
happens over National Knowledge Network (NKN) which has a peak
data bandwidth of upto $1$~Gbps during commissioning
of the HEL1OS operations in the year $2023$.

\subsection{Automated Data Pipeline}\label{sec:automat}

\begin{figure}[!h]
    \begin{center}
        \includegraphics[width=1.0\columnwidth, viewport=196 285 900 645,clip]{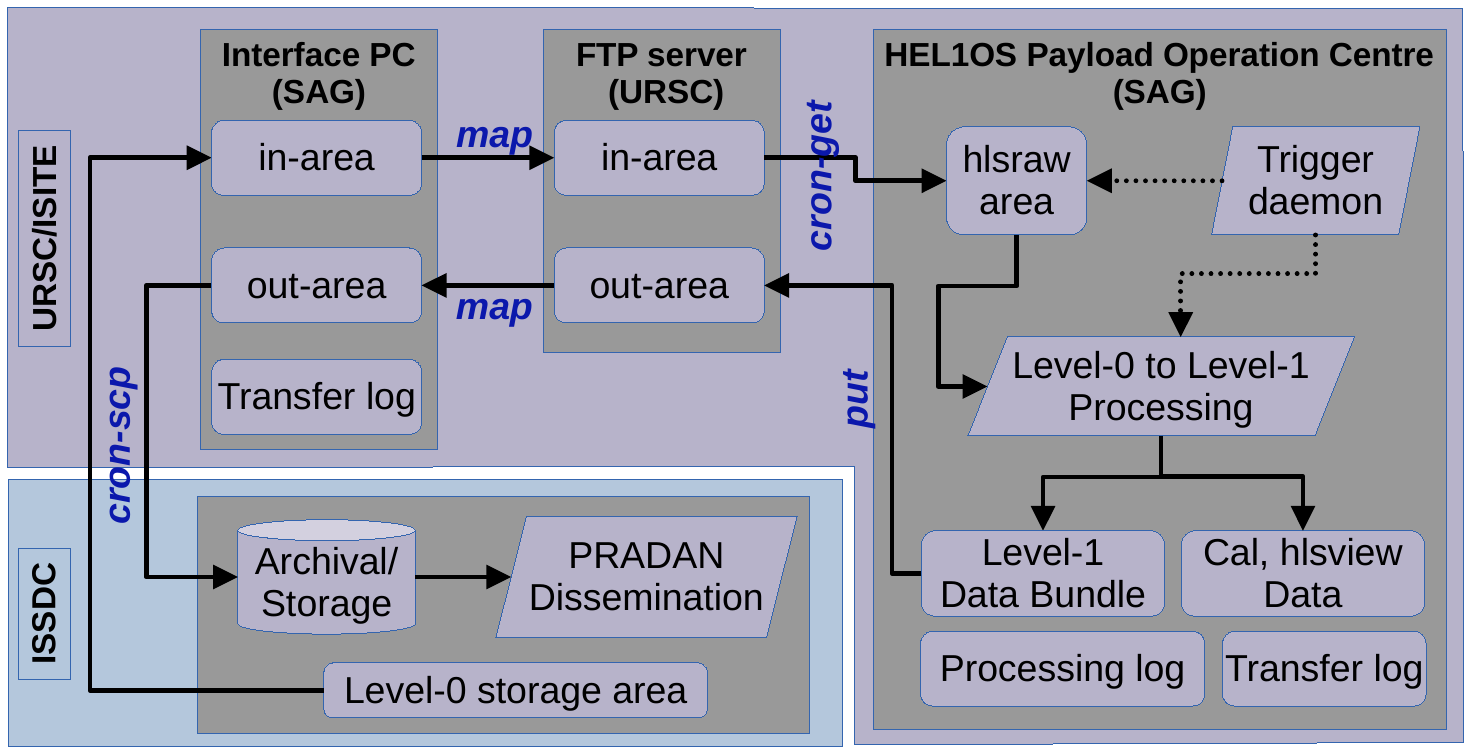}
        \caption{The overall automated data pipeline architecture
        involving the data flow between ISSDC and POC, via an
        interface PC and and FTP server for mapping across networks.
        The automation is achieved by a few concurrent cron jobs and
        daemons. The HEL1OS POC is located at the ISITE campus of
        U.R. Rao Satellite Centre (URSC).}
        \label{fig:overallarch}
        \vspace{-8mm}
    \end{center}
\end{figure}

This whole processing chain of Level-0 generation till dissemination
of the Level-1 data products happens in an automated fashion
(see Figure~\ref{fig:overallarch}).
The data transfer between the data centre at ISSDC and the POC system
is via an interface PC using the secure copy protocol (SCP)
over secure shell (SSH) and File Transfer Protocol (FTP) server.
The data flow across each of these systems are independently automated
involving either polling based or cronjob based periodic data
transfers. For the automation of the transfer dialog,
\textit{TCL/Tk Expect} (\cite{libes1995exploring}) is employed.
These \textit{Expect} scripts triggered by cronjobs automate the
SSH/SCP and FTP sessions while interacting with the respective
servers. The scripts are designed to address possible
connectivity issues to avoid multiple transfer sessions, and also
they maintain detailed logs at each stage to be able to trace upon
any transfer failures. Also, network transmission integrity is
checked via the MD5SUM tags. The trigger daemon running at the POC
system is designed based on the \textit{inotify} tool
(\cite{Love:2005:KKI}). An overseeing \textit{bash} shell-script
takes care of version-numbering of the Level-1 products, ensures
the requisite data archive structure, and takes care of packing
appropriate preview image to be used with the product as its
thumbnail on PRADAN.

\subsubsection{Turn-around time}
The data gets transferred across an interface PC and an FTP server
over a data-diode between ISSDC and POC system, and the turn
around time includes the transfer time across these different
stages as well, other than the Level-0 data processing duration.
Normally two segments of data per day, each of
duration $\sim$~$12$~hr, are dumped at the ground station. These raw
data are processed at ISSDC to generate corresponding Level-0 data and
are made available at a data server. The typical turn around time after
receipt of the Level-0 data at the POC system is
$\sim\mathrm{T_0 + 1\:hr}$. Here $\mathrm{T_0}$ is the time at which
the data polling software run by the POC system transfers
the newly posted Level-0 data from the data server at ISSDC to the
POC system. The instrument operates in the event mode wherein
every photon recorded on the four detectors are saved as part of the
raw-data. The incident event rate increases as a function of
intensity of the solar activity affecting the data volume, data
transfer time, and also the event list processing time. Therefore the
turn around time is longer if there are X-class flares in the data
processed.

\section{Data Pipeline}\label{sec:lev1proc}
\begin{figure}[!th]
    \centering
    \includegraphics[width=0.95\columnwidth, viewport=13 83 581 789,clip]{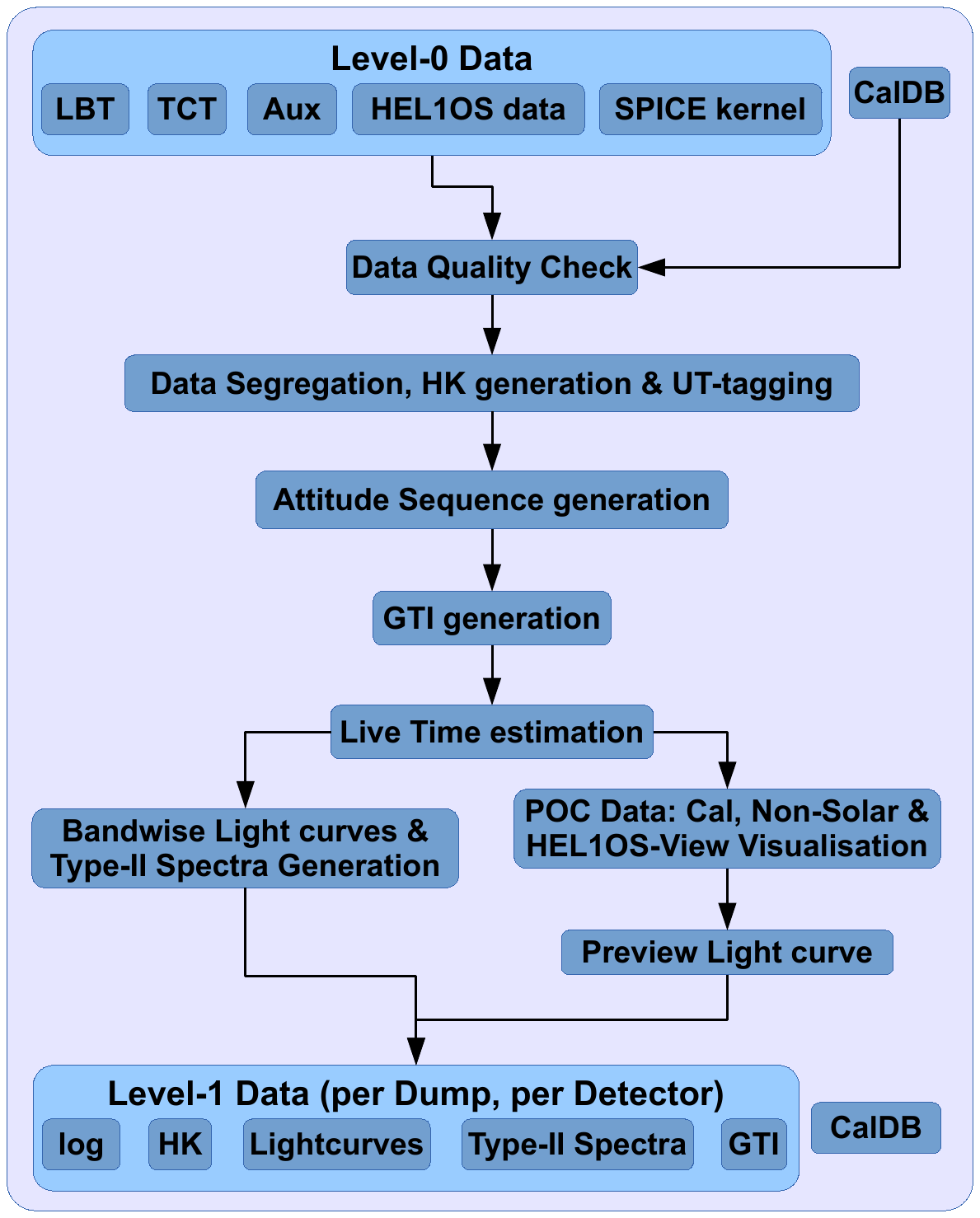}
    \caption{The Level-0 to Level-1 data processing chain run at the POC
    system for each Level-0 data received.}
    \label{fig:lev0-lev1}
    \vspace{-5mm}
\end{figure}

The different stages that are part of the Level-0 to Level-1 data
processing undertaken at the POC system for each Level-0 datafile
received at the POC are depicted in the flow chart in
Figure~\ref{fig:lev0-lev1}.
This Level-0 to Level-1 processing has been developed on
Python~$3.10.9$, and on the production system it is run with
Python~$3.11.5$. Among the other major Python packages which are
employed during this processing include: astropy-$6.1.4$,
matplotlib-$3.7.0$, numpy-$1.23.5$ and spiceypy-$6.0.0$.
At each of the processing stages, detailed log is maintained, and
where ever possible the numpy's vectorised operation capabilities are
employed for improving the processing time. All internal data are
saved in structured arrays.

\subsection{Input to the Processing}
The HEL1OS rawdata is stored on-board in $2048$~bytes sized packets.
Each of these consist of a page sync code at the first byte location,
followed by the House Keeping (HK) data and events data from any of
the four detectors. At the end a $2$~byte Cyclic Redundancy Check
(CRC) value computed for the previous $2046$~bytes is saved. Each
of these packets are also padded with some metadata of the data
handling (DH) system on-board as well as a few other parameters at
the rawdata processing stage making each of them $2120$~bytes long
packets, referred to as 2k~pages. The series of such packets
is what is referred to as `HEL1OS data' in the very first block of
the flow chart. Packed along with this in the Level-0 data are
Low-Bitrate Telemetry (LBT) parameters, the Time Correlation Table
(TCT) file, SPICE kernels and some other auxiliary meta-info.
All these files along with a few set of Calibration Database
(CalDB) files maintained at the POC form the input to the Level-0 to
Level-1 processing. The CalDB files used at this stage include:
(a) list of CZT pixels which are to be ignored (as, either their nominal
background level is too high, or, they are some of the edge pixels
whose response is not determined), (b) different energy sub-bands and the
full-band for which the products need to be generated, and,
(c)  the
energy-channel relationship in terms of temperature dependent gain
and offset of the four detectors. Currently, data from $47$~pixels of
the CZT1 detector and from $19$~pixels of the CZT2 detector
among $256$~pixels in each, are being ignored.

The auxiliary data packed with the payload data within the Level-0
data which are used in the processing are: the TCT and the SPICE
Camera Matrix Kernel -- both in binary format. The TCT is a
correlation of (a) the on-board time (OBT) of the DH system
(DH-OBT), (b) that of the on-board computer (OBC), and (c) the
corresponding Coordinated Universal Time (UTC) determined as part
of the ground processing.  The HK parameters including both analog
values and status bits of different instrument telemetry parameters
are also packed in the Level-0 data in ASCII format.

As part of the very first stage of `data quality check', the page sync
codes and the CRC codes are checked for in each 2k~page of the
`HEL1OS data'. All the frames which fail either of these checks are
dropped from further analysis. The sequence numbers of all clean
2k~pages which pass these checks are noted.

\subsection{Data Segregation, HK generation and UT-tagging}
\subsubsection{HK Data handling}
In these clean 2k~pages, the metadata read from the page envelope
are: frame sequence numbers assigned in the DH and
at the rawdata processing stage, ground receive local time,
DH-OBT and UTC corresponding to it. Also the HK parameters saved
on-board by the instrument are read, such as: recorded temperature
values of each of the four detectors, status-bits of different
onboard logics, page start time, count rates of each detector,
energy threshold values of each detector, peaking time, latest hot
pixel, their counts, hot pixel threshold of both CZT detectors,
HV analog monitor values of both CZT and CdTe detectors, the
pile-up counter values of the two CdTe detectors, saturation counters
and bunched pixel counters of the two CZT detectors. Both metadata
in the envelope as well as the HK parameters which are present at
fixed byte locations of the 2k~pages, are read and processed in one
go exploiting the numpy's vectorised operation. The HK parameters are
also ultimately saved as part of the Level-1 product bundle that is
disseminated, and some of them are used for generation of good time
intervals (GTI) as well.

\subsubsection{Event data handling}
The vast section of consecutive $2004$~byte locations in every 2k~page
is meant for the events registered in any of the four detectors and it
consists of information saved for every photon recorded as and when
the electronics system registers them.
The information saved for a given event consists of $2$~bits for the
detector ID (to mark one of the four detectors), $9$~bits of
Analog to Digital (ADC) channel number and $5$ least significant
bits of the HEL1OS clock. If the event is registered in one of the CZT
detectors, along with the above $8$~bits of the pixel-ID is also
saved. Therefore information about a CdTe event consists of $16$~bits
and a $24$~bits for a CZT event.
For each of the events we also save the latest temperature values
and the record numbers from respective HK section.
Only the $5$ least significant bits of the clock are saved for each
event as mentioned above to conserve on board memory. The
complete HEL1OS-OBT for each event is constructed by using the $20$
most significant bits from the page start time of the respective
2k~page, after accounting for possible overflow in these $5$ bits
which depends on the page-filling rate. 

\subsubsection{Time correlation scheme}
The DH-OBT recorded on a given page is offset from the page start
time of the subsequent page by $2.2$~$\mu$s which is much smaller
than the instrument's clock resolution of $10$~ms. These two time
sample pairs of HEL1OS page start time and the latched DH-OBT are
used along with the TCT entries made available, for the time
correlation exercise to compute the UTC corresponding to each page
start time. With the $25$~bits saved for the page start time and
the clock resolution of $10$~ms, the page start time will overflow
in every $\sim$~$93.2$~hr. In order to handle possible overflow of
these $25$~bits clock within a data dump, its bitshifted value
along with the sequence number of the corresponding 2k~page is
used for all internal referencing. 
The UTC values obtained are saved in the units of Modified Julian
Date (MJD). In addition, the time sample pairs of the DH-OBT and the
page start time are ensured to be from consecutive 2k~pages, 
after filtering repeating pair values (which happens when the page
filling rate is high for enhanced solar activity).

The MJD values of every event recorded in the four detectors read
from all the 2k~pages are computed by interpolating the time-series
of the page start time values from every 2k~page and their
corresponding MJD values. Subsequently, the ADC channel numbers of
each event are converted to corresponding energy values using the
temperature dependent channel-energy relationships pre-computed
during ground calibration exercise. Each of these are undertaken
in numpy vectorised calls per detector data. During the PV phase,
the calibration parameters obtained from ground experiments have
been verified using onboard radioactive source spectra from
quiet-time and blank sky observations.

\subsection{Attitude Sequence Generation}
In the SPICE kernel handling module, a time series with following
aspect parameters are determined at $1$~s cadence for the observation
time window: Right Ascension of Sun, Declination of Sun, Flag to
indicate if the Sun is within the field of view of the instrument,
separation between Sun vector and +Yaw spacecraft body axis,
separation between Sun vector and +Roll spacecraft body axis,
separation between Sun vector and +Pitch spacecraft body axis,
Right Ascension of +Yaw axis and Declination of +Yaw axis. These are
part of the `Attitude Sequence generation' marked in the flowchart in
Figure~\ref{fig:lev0-lev1}. For this exercise following SPICE kernels
in ASCII format are employed: clock cooefficients SCLK, Frame Kernel,
Instrument Kernel, Leap-seconds Kernel, and Planetary Constants
Kernel. Along with these, following binary kernels are also used:
Spacecraft Ephemerides Kernel, Planetary Ephemerides Kernel and
Attitude (or `camera-matrix') Kernel. The last two kernels vary for
each observation and are therefore computed and packed as part of
every Level-0 data bundle. The remaining kernels are static and are
maintained at the POC system. The aspect angles generated are also
made available in the HK product saved in the Level-1 bundle.

\subsection{GTI Filtering}
The `GTI generation' mentioned in the flow chart involves GTI
durations based on different parameters. The GTI parameters common
among all the four detectors are the frame-drops and whether the Sun
was within the instrument FOV or not. The respective HV value are
common among the CdTe detector pairs, and the CZT detector pairs.
The detector specific GTI parameter is the on-board measured
temperature values for each. However, in the nominal operating
configuration the temperature-read module of the two CZT detectors
are disabled (as it affects the events-recording). Instead the
recorded HV values of the CZT detectors themselves are considered
proxy for their temperature values, for the reason that the onboard
safety logic would disable the CZT HV whenever the onboard
detector-tray temperature goes off preset limits. Ultimately two sets
of GTI files for each of the four detetors are generated -- one set
for the Sun-pointing intervals and another set for the pointings away
from the Sun. These GTI files are made available as part of the
Level-1 bundle disseminated as well.

Based on the generated GTI intervals specific to each of the four
detectors, the respective event-lists are filtered. These clean event
lists, complying with the OGIP FITS format, also form one of the data
products in the Level-1 bundle that is disseminated.

\subsection{Product Generation}\label{sec:proddetails}
The clean event lists are used to generate light curves in
pre-specified energy sub-bands (listed in the user manual) at a
default cadence of $1$~s saved in different binary table extensions
for each detector.
The FITS Type-II Spectra are also generated using these clean event
lists for each detector at a default time cadence of $20$~s.
Currently generated Spectral Redistribution Function (SRF) for the
two detectors are of following energy range and spectral cadence:
CdTe detectors -- $1.88$~keV to $89.98$~keV over $511$ channels
corresponding to $\sim$~$0.172$~keV per channel, and CZT detectors
-- $18$~keV to $160$~keV over $341$~channels corresponding to
$\sim$~$0.416$~keV per channel. The Type-II spectra are generated
for these parameters.

\begin{figure}[!h]
    \centering
    \includegraphics[width=\columnwidth, viewport=41 241 551 603,clip]{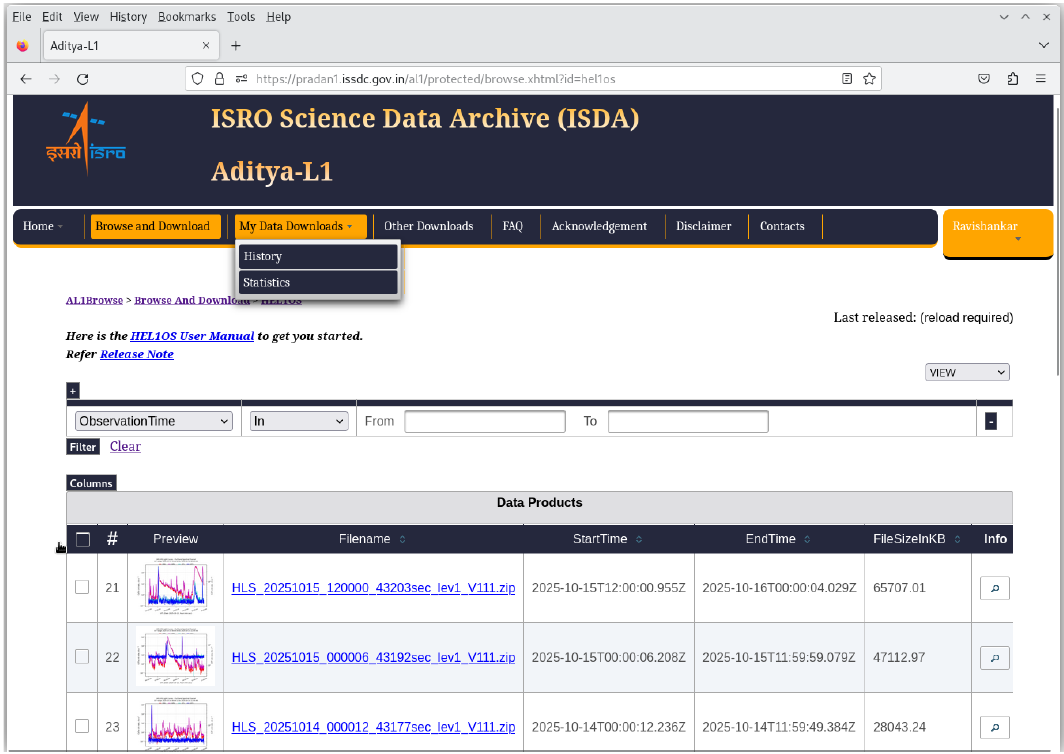}
    \caption{The HEL1OS page on PRADAN portal of ISSDC with the
    `Preview' thumbnail images for each listed product, search
    dialogue and the links to the user manual and release notes; at
    the bottom of this webpage (not shown here), is a link anchored
    to the HEL1OS section of `Other Downloads' page of PRADAN.}
    \label{fig:pradan}
\end{figure}

All pointings which are away from Sun, including the planned
observations of the stellar calibration source (such as Crab) are
flagged and not disseminated. There is also `hlsview' visualisation
module used as part of the pipeline in the POC to generate quicklook
plots of lightcurve at $1$~min cadence, coarse spectra, spectrograms
of all the four detectors and the pixel distribution of the two CZT
detectors -- all these for the entire playback duration for ready
health check of the instrument. The coarse lightcurve plot generated
by this visualisation module with the lightcurves of all the four
detectors is also passed on with the Level-1 data bundle to be used
as a `thumbnail preview image' on the PRADAN dissemination page
(see Figure~\ref{fig:pradan}). The structure of the Level-1 data
bundle and description are provided in the user manual.

\vspace{-6mm}
\section{Data Product Dissemination}\label{sec:dissem}

The Level-1 data products generated at the POC are transferred to
ISSDC data server for dissemination. Along with the standard products
of HEL1OS, the Level-1 transfer bundle also consists of an Extensible
Markup Language (XML) with relevant metadata required at the PRADAN
user interface.

The Level-1 data product bundles transferred to the ISSDC data server
are ingested and made available through the PRADAN portal. The URL of
the portal is \verb+https://pradan1.issdc.gov.in/al1/+,
and the data is accessible to all the registered users. 
HEL1OS page on PRADAN is shown in Figure~\ref{fig:pradan} in which
each Level-1 data bundle is listed along with the start and end times
of the observation (in UTC) and also the preview thumbnail which is
also sent along with the Level-1 data bundle. The `Info' button for
each row when clicked lists all the associated metadata.

Also present on the HEL1OS PRADAN page are hyperlinks to the user
manual and release notes at the top. At the bottom is present a
hyperlink anchored to the HEL1OS section of the `Other Downloads'
page. The current listing in this section consists of:
{\footnotesize
\begin{verbatim}
HEL1OS_UserManual.pdf
CAL_epoch20231001_CdTeResponseReader.zip
CAL_epoch20231001_CZTResponseReader.zip
hel1osLightCurvePlotDisplay_Ver1.2.py
HEL1OS_ReleaseNotes.pdf 
\end{verbatim}
}

The two zip-files mentioned above consist of the response files of the
CdTe and CZT detectors (currently for the epoch starting at $20231001$),
and also SolarSoft Object Spectral Executive (OSPEX) reader code in
Interactive Data Language (IDL) for the two detectors. More details
of their usage in the data analysis, and also of the Python utility
for the HEL1OS lightcurves are present in the user manual. The future
releases of new utilities as well as the updates made to the
instrument response, user manual, and other information useful for
data analysis, will be listed on this `Other Downloads' section of
the PRADAN portal.

\section{View of Sun in hard X-rays}\label{sec:hardsun}

\subsection{Hard X-ray detections of flaring events}

\begin{figure*}[!h]
    \centering
    \includegraphics[width=0.86\textwidth]{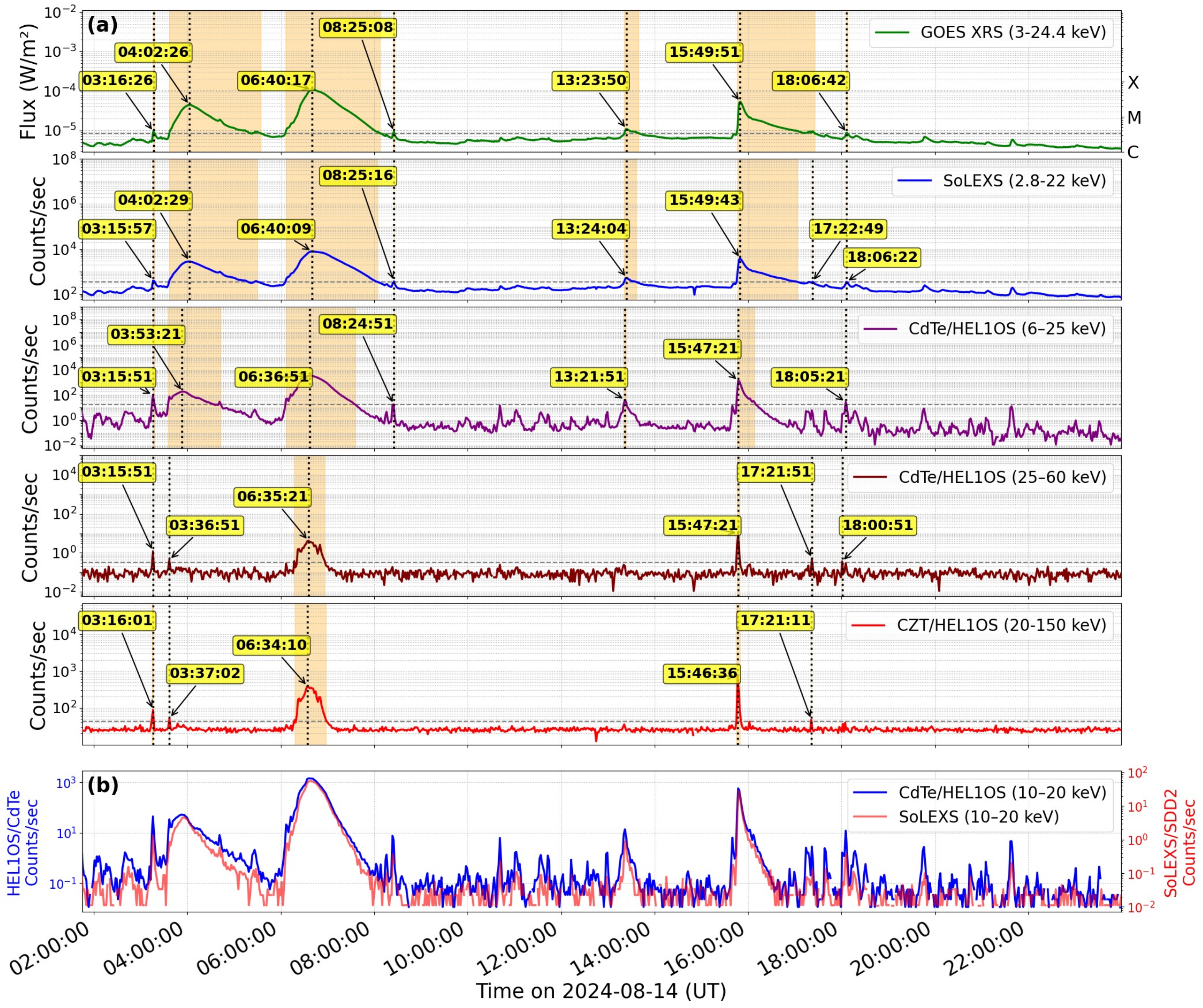}
    \caption{(a) Light curves of GOES XRSA, SoLEXS, along with those of
    the CdTe and CZT  detectors of HEL1OS for August $14$, $2024$.
    The vertical dashed lines and the UTC values marked are at the
    detected peak-flux of different detected flares. The shaded
    regions are marked between the start and end times of each
    flare. The horizontal dashed line is set at 
    detection significance level for each
    detector. This is based on both computed background $\sigma_{b}$
    and chosen $\sigma$-levels $N$, for detection of the weakest among
    GOES catalog flares for that day -- C8.6 class, in each case. The
    background $\sigma_{b}$ values are computed for prevailing
    quiescent flux during $21$:$45–23$:$59$~UTC for each case.
    The peak-time values across different instrument lightcurves are
    different owing to the different energy-bands as indicated. In the
    GOES lightcurve plot, the C, M and X class flux levels are marked
    on the second Y-axis. More details of the detected flares are
    present in Table~\ref{tab:lccross}. (b) Comparing
    the data from SoLEXS and CdTe/HEL1OS for same energy band
    of $10-20$~keV over the same period on August $14$, $2024$. The
    temporal profile of the solar activity in both the detectors are
    in full agreement.}
    \label{fig:lccross}
\end{figure*}

\begin{table*}[!h]
    \centering
    {\normalsize
    \caption{Details of the peak-times (UTC in
    HH:MM:SS) of the flares from GOES Catalog on
    August $14$, $2024$, their detections by SoLEXS and HEL1OS' CdTe
    \& CZT detectors;
    also indicated are: the background $\sigma_{b}$
    values and $\sigma$-level $N$, set for each case (refer to
    Figure~\ref{fig:lccross}(a) and its caption).}
    \label{tab:lccross}
    \begin{tabular}{|C{8mm}|C{2.9cm}|C{2.9cm}|C{2.9cm}|C{3.1cm}|C{2.9cm}|}\hline
        GOES Class & \
        GOES (${3-24.4}$~keV) \small $\sigma_{b}$~$\sim$~$2.8e-07$~Wm$^{-2}$\newline N = 30 & \
        SoLEXS (${2.8-22}$~keV) \small $\sigma_{b}$~$\sim$~$11.7$~counts s$^{-1}$\newline N = 30& \
        CdTe (${6-25}$~keV) \small $\sigma_{b}$~$\sim$$~0.17$~counts s$^{-1}$\newline N = 100 & \
        CdTe (${25-60}$~keV) \small $\sigma_{b}$~$\sim$~$0.033$~counts s$^{-1}$\newline N = 10 & \
        CZT (${20-150}$~keV) \small $\sigma_{b}$~$\sim$~$2.16$~counts s$^{-1}$\newline N = 20 \\\hline
        C$9.2$ & $03$:$16$:$26$ & $03$:$15$:$57$ & $03$:$15$:$51$ & $03$:$15$:$51$ & $03$:$16$:$01$ \\\hline
        M$4.4$ & $04$:$02$:$26$ & $04$:$02$:$29$ & $03$:$53$:$21$ & $03$:$36$:$51$ & $03$:$37$:$02$ \\\hline
        X$1.1$ & $06$:$40$:$17$ & $06$:$40$:$09$ & $06$:$36$:$51$ & $06$:$35$:$21$ & $06$:$34$:$10$ \\\hline
        M$1.0$ & $13$:$23$:$50$ & $13$:$24$:$04$ & $13$:$21$:$51$ & - & - \\\hline
        M$5.3$ & $15$:$49$:$51$ & $15$:$49$:$43$ & $15$:$47$:$21$ & $15$:$47$:$21$ & $15$:$46$:$36$ \\\hline
        C$8.6$ & $18$:$06$:$42$ & $18$:$06$:$22$ & $18$:$05$:$21$ & $18$:$00$:$51$ & - \\\hline

    \end{tabular}
    } 
    \vspace{-3mm}
\end{table*}

The two X-ray payloads on Aditya-L1 mission itself, SoLEXS
(\cite{Sankar:SolPhy}) and HEL1OS (\cite{Anuj:solphy}) provide a
continuous and complementary view of the Sun spanning a broad
energy range of $\sim$~$2.8$~keV to $150$~keV. This enables
investigation of the temporal evolution of solar flare emission
across both relatively soft and hard X-ray energy bands. 
Shown in Figure~\ref{fig:lccross}(a) are the light curves plotted
for August $14$, $2024$ with the data from GOES XRSA
($0.5-4.0$~\AA~or $3.1-24.8$~keV), SoLEXS ($2.8-22$~keV),
CdTe/HEL1OS ($6-25$~keV), CdTe/HEL1OS ($25-60$~keV) and CZT/HEL1OS
($20-150$~keV) plotted in different sub-panels in this order, from
top to bottom, to illustrate the X-ray temporal profile of the
solar variability as viewed in different X-ray energy bands. Shown
in Figure~\ref{fig:lccross}(b) is the plot comparing the lightcurves
of SoLEXS and CdTe/HEL1OS in a common energy-band of $10-20$~keV for
the same time interval on August $14$, $2024$. The data from the two
detectors exhibit closely matching temporal profiles of the solar
activity in this energy band. The difference in the count-rates of
the two is due to the difference in their effective area.

The peak detection method applied on the light curves
in Figure~\ref{fig:lccross}(a) involved dynamic thresholding. For
this, standard deviation of the background $\sigma_{b}$, was computed
over the time interval $21$:$45–23$:$59$~UTC which is the nearest
period noticed without significant solar X-ray activity. The peak
threshold was set at $\sigma$-level ($N$) times $\sigma_{b}$, wherein
$N$ was set at a suitable value to be able to pick even the weakest
flare in the GOES catalog, in the selected data interval -- C8.6 class
(at the GOES peak time of $18$:$06$:$42$ marked in the Figure).
These detector-specific $\sigma_{b},N$ value-pairs are mentioned in
Table~\ref{tab:lccross}. It is to be noted that the high peak
significance values of $N$ in all these cases are due to the high
signal to noise ratio of the data spanning a few orders, evident in
Figure~\ref{fig:lccross}.

The UTC values in the text-boxes in Figure~\ref{fig:lccross}(a)
indicate the time values of detected peaks. The CdTe and CZT data used
to make this plot are combined from both the respective detector pairs.
The CdTe data were subdivided into two energy ranges, $6-25$~keV
and $25-60$~keV, to examine the spectral hardness of the detected
flares and to enable direct comparison across instruments operating
in overlapping energy domains. As expected, essentially all flares
detected in the softer energy bands of GOES and SoLEXS are also
visible in the lower-energy CdTe band ($6–25$~keV), since this
energy range contains a substantial contribution from thermal plasma
heated during flare energy release. However, a few of these events
(GOES M$1.0$ class peaking at $13$:$23$:$50$ UTC and GOES C$8.6$
class peaking at $18$:$06$:$42$ UTC listed in Table~\ref{tab:lccross})
do not produce detectable emission in the higher-energy CdTe
($25–60$~keV) or CZT ($20–150$~keV) bands, indicating that the
emission at higher photon energies is either weak or absent for
those events.
CdTe/HEL1OS light curves also show flare-like enhancements that
are more clearly visible in its $6-25$~keV energy bands and which
are not observed in the softer energy channels of the other
instruments (i.e., GOES XRSA ($3.1-24.8$~keV) and SoLEXS
($2.8-22$~keV)). This is of course dependent on the peak-detection
levels set and the detection methods adopted as well.
For instance, the peak detection method adopted
here has picked an uncatalogued flux enhancement in GOES data at
$08$:$25$:$08$ UTC (see Figure~\ref{fig:lccross}(a)). The method has
also picked corresponding detections in SoLEXS and CdTe $6-25$~keV
data for this enhancement. Likewise the flux enhancement picked by the
method at $17$:$22$:$49$ UTC in SoLEXS data (with corresponding
detections in CdTe and CZT data), is neither present in the GOES
catalog, nor picked up by the method in the GOES data.

\begin{figure*}[!h]
    \centering
    \begin{minipage}[t]{\textwidth}
        \centering
        \includegraphics[width=0.9\textwidth]{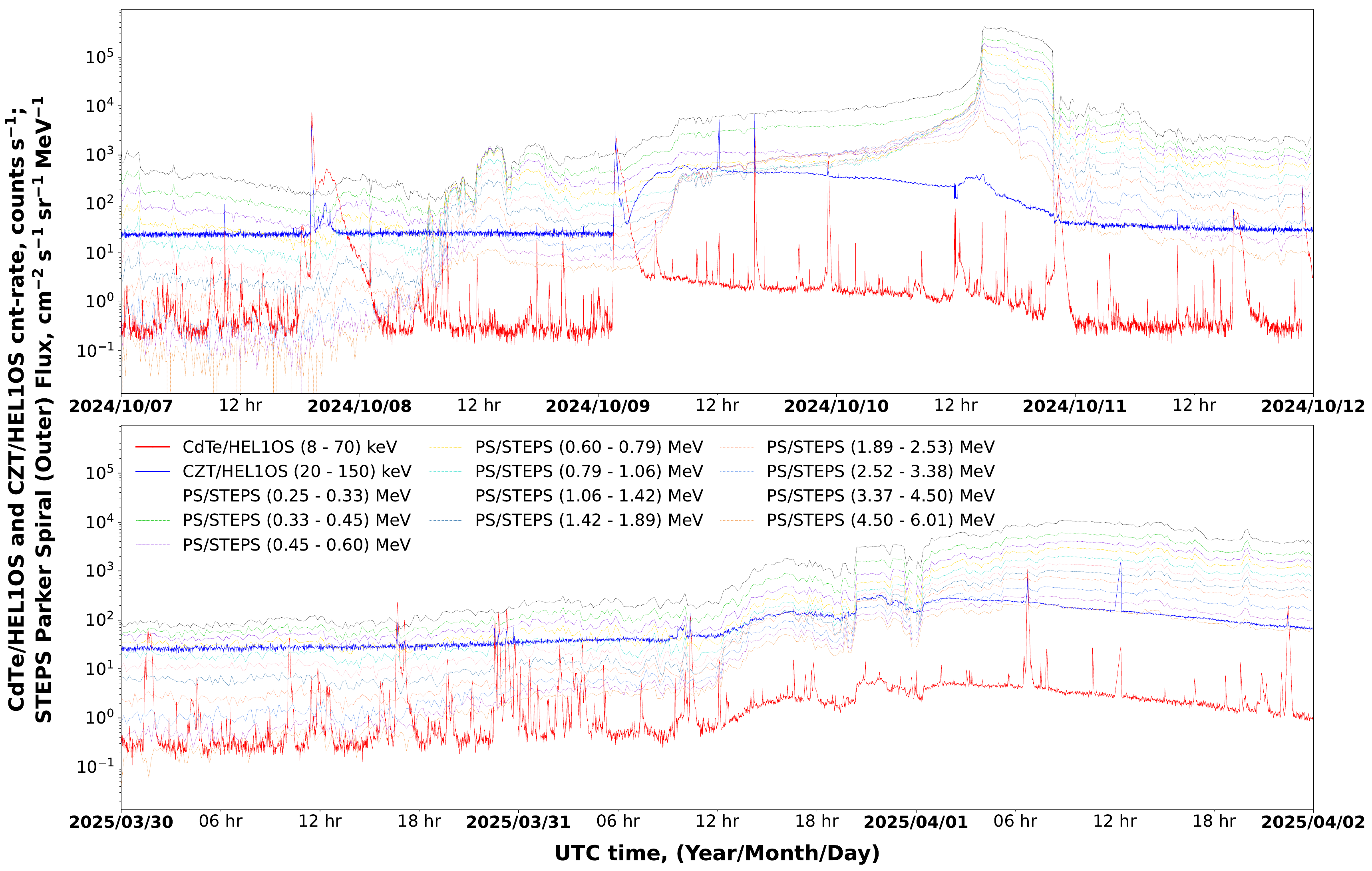}
        \caption{Comparing the lightcurves of HEL1OS' CZT (blue;
        $20-150$~keV) and CdTe (red; $8-70$~keV) detectors with those
        of the ASPEX/STEPS-PS detectors in different bands (in other
        colours) for two time intervals 
        around $2024$~October~$09$ (top panel) and around $2025$~March~$31$
        (bottom panel). This is to illustrate the SEP-driven
        background variations in the HEL1OS detectors. The legend is
        shown only in the bottom panel for lack of space, but same
        applies to the top panel plot as well.}
        \label{fig:hls_stps}
    \end{minipage}
    \vspace{-5mm}
\end{figure*}

An important and expected observational characteristic is that the
peak times of the flare emission differ systematically across
energy bands. Specifically, the harder X-ray emission detected by
HEL1OS generally peaks earlier than the softer X-ray emission
detected by GOES and SoLEXS. For events detected in both HEL1OS
detectors, the peak in the higher-energy CZT band often precedes
that in the lower-energy CdTe band.
These observed characteristics can be understood in the framework
of the standard model of solar flares and the Neupert effect (for
example, \cite{1968ApJ...153L..59N}; \cite{1993SoPh..146..177D};
\cite{2002A&A...392..699V}).
In this picture, hard X-ray emission, resulting from accelerated
electrons interacting with dense plasma, signals the direct energy
release and particle acceleration during the flare impulsive phase.
These electrons also heat the chromosphere, causing plasma expansion
and subsequent delayed thermal soft X-ray emission. The observed
earlier peaks in harder HEL1OS bands and delayed peaks in softer GOES,
SoLEXS, and lower-energy CdTe channels support this sequence. Flares
with strong hard X-ray detection indicate efficient particle
acceleration and significant non-thermal emission, while the absence
of higher-energy emission suggests insufficient high-energy electron
populations (see Table~\ref{tab:flarelist} for a list of HEL1OS Hard
X-ray flares; details will be presented in \cite{Manju:inpreparation}).
Thus, multi-energy observations from HEL1OS, GOES, and SoLEXS are
vital for diagnosing flare energy release, particle acceleration,
and plasma heating.

\subsection{Background level}\label{sec:bg}

The background level of the HEL1OS detectors has been stable ever
since it was switched ON. This can be both due to the shielding offered
by the satellite body to the detectors, and also due to the benign
orbit of Aditya-L1 compared to the inner \& outer radiation belts and
South Atlantic Anamoly regions if it were to be a low-Earth orbit.
Typically, a background level of
$\sim$~$0.1$~$\mathrm{counts\:s^{-1}}$ for CdTe detectors and
$\sim$~$20$~$\mathrm{counts\:s^{-1}}$ for CZT detectors has been
observed in their respective overall operating energy bands. The
enhancements in the background are solely driven by the solar activity
with increased SEP flux.
Shown in Figure~\ref{fig:hls_stps} are the lightcurves of the HEL1OS
detectors (combined for the respective pair, and filtered for the
background levels mentioned above), along with the differential
lightcurve of the ASPEX/STEPS-PS detectors, for two date ranges of
$2024$~October~$07$ -- $2024$~October~$11$
(on the top panel) and $2025$~March~$30$ -- $2025$~April~$01$ (on the
bottom panel). In both instances, the simultaneity of the background
level variations indicate the prevalent solar particle activity
(seen clearly in the bottom panel from $\sim\:12$~hr UTC on
$2025$~March~$31$). In addition, the top panel also depicts instances
of the lag in the lightcurves between the two instruments, by a few
hours, the duration expected of the particle travel time from Sun
till the L1 point (for example around $\sim\:6$~hr UTC on
$2024$~October~$09$ in the top panel).
For details on the generation of the HEL1OS SEP Contamination
Database, please refer to \cite{Unnimaya:inpreparation}.

\subsection{Long term hard X-ray lightcurve}

\begin{figure*}[!h]
    \centering
    \includegraphics[width=0.98\textwidth]{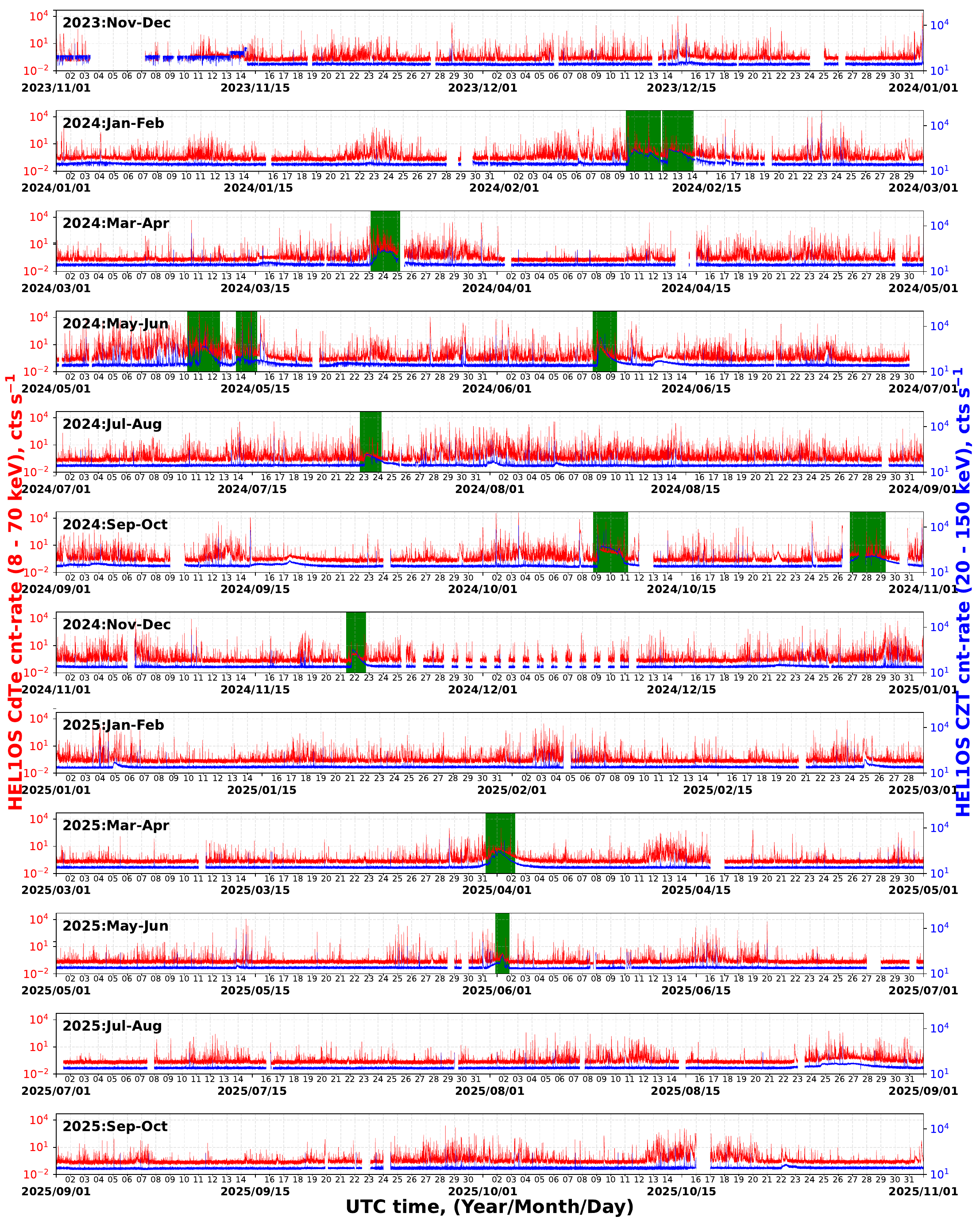}
    \caption{Light curves using the combined data from the CdTe (red;
    $8-70$~keV) and CZT (blue; $20-150$~keV) detectors from
    $01$~November,
    $2023$, a few days after HEL1OS was commissioned till $31$~October,
    $2025$, two months in each row. Each of the excursions of the flux
    from the background level indicate solar flare activity, some of
    which are hard enough to be detected by CZT detectors as well
    (listed in Table~\ref{tab:flarelist}).
    The gaps in the light curves are due to data non-availability.
    The green shaded regions indicate the intervals which lasted for
    a few days with higher CZT background due to enhanced solar
    activity.}
    \label{fig:lclong}
    \vspace{-5mm}
\end{figure*}

HEL1OS provides a continuous view of the Sun in hard X-rays from the
vantage point of \textit{Aditya-L1} mission's orbit about the
Lagrangian Point L1. This is alongside the monitoring undertaken by
Spectrometer/Telescope for Imaging X-rays (STIX) on the Solar Orbiter,
Hard X-ray Imager (HXI) on Advanced Space-based Solar Observatory
(ASO-S) as well as Konus-Wind, which are the other currently active
hard X-ray instruments monitoring Sun.

Plotted in Figure~\ref{fig:lclong} are the long term light curves of
the CdTe and CZT data, from $01$~November, $2023$ till
$31$~October, $2025$, with data for two months at a
time in each row of the plot. The data is combined for the respective
detector pair as mentioned earlier, for the complete energy bands of
each. The $8-70$~keV CdTe data is plotted in red (on the left
side Y1-axis) and the $20-150$~keV CZT data in blue (on the right side
Y2-axis). The gaps in the plots are
either due to operational reasons, data non-availability \& issues,
or intervals when Sun did not appear in the FOV of HEL1OS. The
apparent shift in the background level in the initial days till mid
November, $2023$ are due to higher thresholds set in the payload.
The temperature read module of CZT was kept ON till Jun, 2025 and
intermittently, the two CZT detectors do not record events in those
intervals when the temperature is being read. Those have been suitably
taken care of while combining the data from both CZT detectors. 
The combined data is also filtered for observed background level as
mentioned earlier.
Each of the flux excursions represent a solar flare activity. At times,
the gradual shift in the background level over days can also be seen
which are associated with enhanced solar activity some of which
corresponds to CME events and associated higher incident SEP events as
well. This has been marked with green shaded regions in
Figure~\ref{fig:lclong} for higher background intervals which lasted
for a few days.
For illustration, the interplanetary shock data base version-2
compiled with Wind, Advanced Composition Explorer (ACE), and Deep
Space Climate Observatory (DSCOVR) observations collected at the
Lagrangian point L1 (\cite{denny_m_oliveira_2025_15121223}), lists
the following UTC of shock observation at L1 for some
of which analogous background enhancements in the HEL1OS detectors
can be seen in Figure~\ref{fig:lclong}:\\
$2024$/$02$/$11$, 
$2024$/$02$/$13$, 
$2024$/$02$/$24$, 
$2024$/$03$/$24$, 
$2024$/$05$/$10$, 
$2024$/$05$/$16$,
$2024$/$06$/$10$, 
$2024$/$07$/$23$, 
$2024$/$10$/$10$,
$2024$/$10$/$26$, 
$2024$/$10$/$28$.

It should also be noted that all the background enhancements in the
$20-150$~keV light curves of the CZT detectors are associated with
those in the $8-70$~keV light curves of the CdTe detectors. There
are however many instances as can be seen in the Figure when these
enhancements have occured in the light curves of the CdTe detectors
only, and no associated changes in the ligth curves of the CZT
detectors are seen.

Over the past two years the transitioning of the current solar cycle
$25$ from its maxima towards minima is also evident from the gradual
decrease in the frequency and intensity of the flaring activities as
can be seen from these long term lightcurves.

Provided in the Table~\ref{tab:flarelist} are the details of
detections of hard X-ray flares with the combined CZT data for
about a little over an year since July, $2024$.
The selection criteria included detection in the energy range of
$40$~keV~$-$~$150$~keV and a minimum of $5000$~events per flare.
Also provided are the fluence values which are the total
integrated net counts per unit detector geometric area
($\text{counts cm}^{-2}$) in the $40$~keV~$-$~$150$~keV range over
the duration of the flare emission in the CZT detectors.

\subsection{Science Data Product}
The science-ready Level-1 products include the Light curves and
Spectra, both as OGIP compliant FITS data files. Both light curves
and spectra are made available for individual detectors, i.e.
separately for CdTe1, CdTe2, CZT1 and CZT2 detectors.
The section~\ref{sec:proddetails} gives a detailed account of the
formats, energy ranges and cadences of the data products which are
disseminated. More details including the directory structure are
part of the user manual made available on the PRADAN portal. In
this section a glimpse of some products are provided which
demonstrate the key science capabilities of HEL1OS.
For this, representative C-class, M-class, and X-class solar flares
duing their impulsive phases are selected for combined CdTe and CZT
data. Here, classes are as classified by the peak flux in the GOES
XRS $1-8$~\AA~channel. Following flares are considered here, to
demonstrate temporal and spectral capabilities of the instrument
over a vast dynamic range of the peak flux: (a) GOES C$6.5$-class
flare that occured on May $31$, $2025$, (b) GOES M$5.0$-class flare
on July~$17$, $2024$, and, (c) GOES X$2.7$-class flare on May $14$,
$2025$.
For each of these flares we have first presented their lightcurves
in Figures~\ref{fig:lcC},~\ref{fig:lcprod} and~\ref{fig:lcX}
respectively. Then we have presented their spectral fits in
Figures~\ref{fig:specC},~\ref{fig:specM} and~\ref{fig:specX}
respectively.
Detailed descriptions of python-based analysis tools for processing
the Level-1 event lists to generate deadtime corrected temporal and
spectral products are discussed in \cite{Manju:inpreparation}.

\subsubsection{Lightcurve Product Analysis}
\begin{figure}[!ht]
    \centering
    \includegraphics[width=\columnwidth, viewport=71 18 832 432,clip]{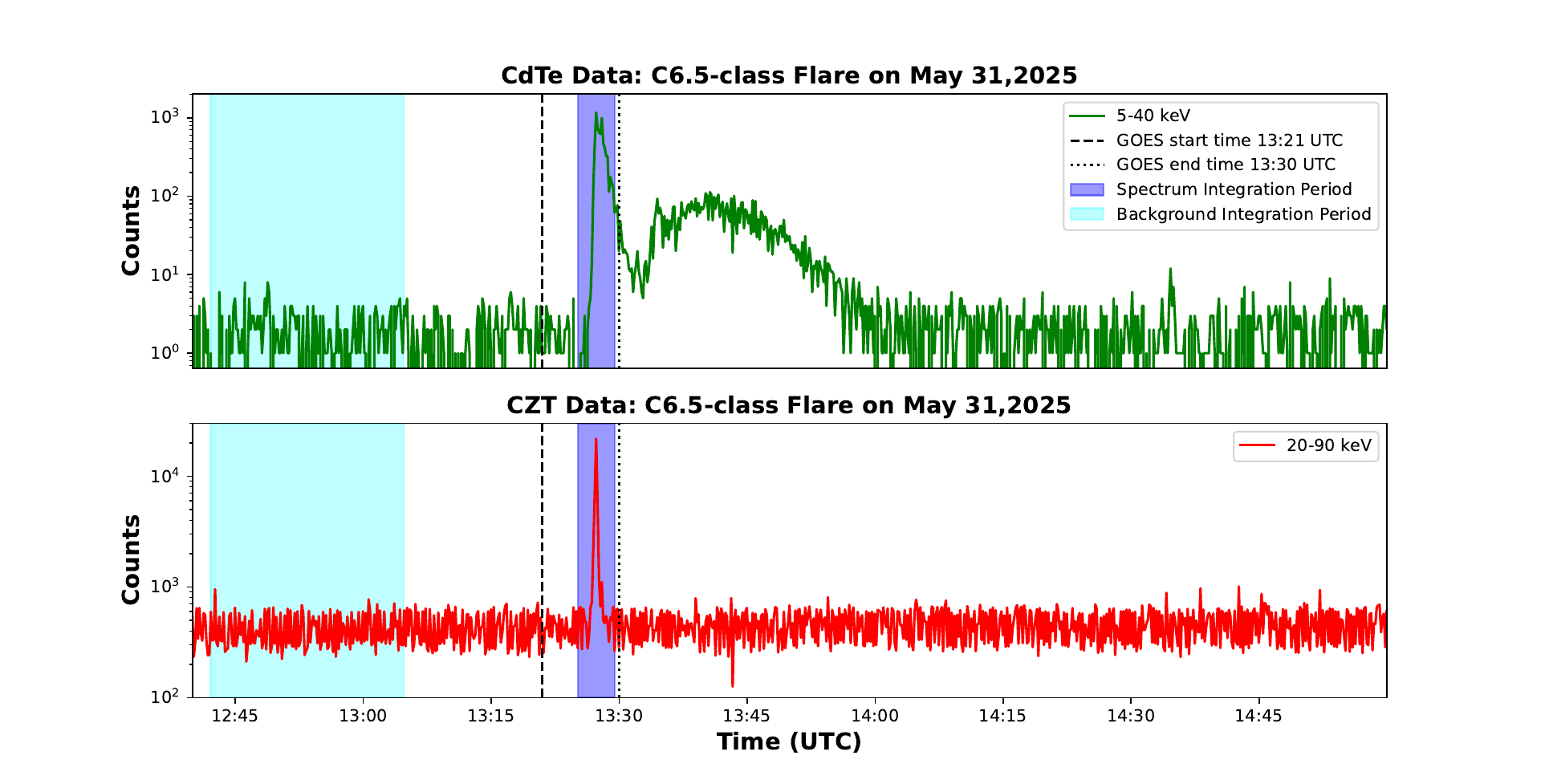}
    \captionof{figure}{Lightcurve of the GOES C$6.5$ class flare that
    occured on May $31$, $2025$. In the top panel is the combined
    CdTe data in $5-40$~keV band, and in the bottom panel is the
    combined CZT data in $20-90$~keV band. The purple bar marked is
    for the interval over which the integrated source spectra in
    Figure~\ref{fig:specC} are obtained (that is $13$:$25$:$09$
    $-$ $13$:$29$:$29$ UTC). Similarly, the cyan bar marked indicates
    the interval over which the background spectra was obtained
    (between $12$:$42$:$07$ $-$ $13$:$04$:$47$ UTC). }
    \label{fig:lcC}
\end{figure}

\begin{figure}[!ht]
    \centering
    \includegraphics[width=\columnwidth]{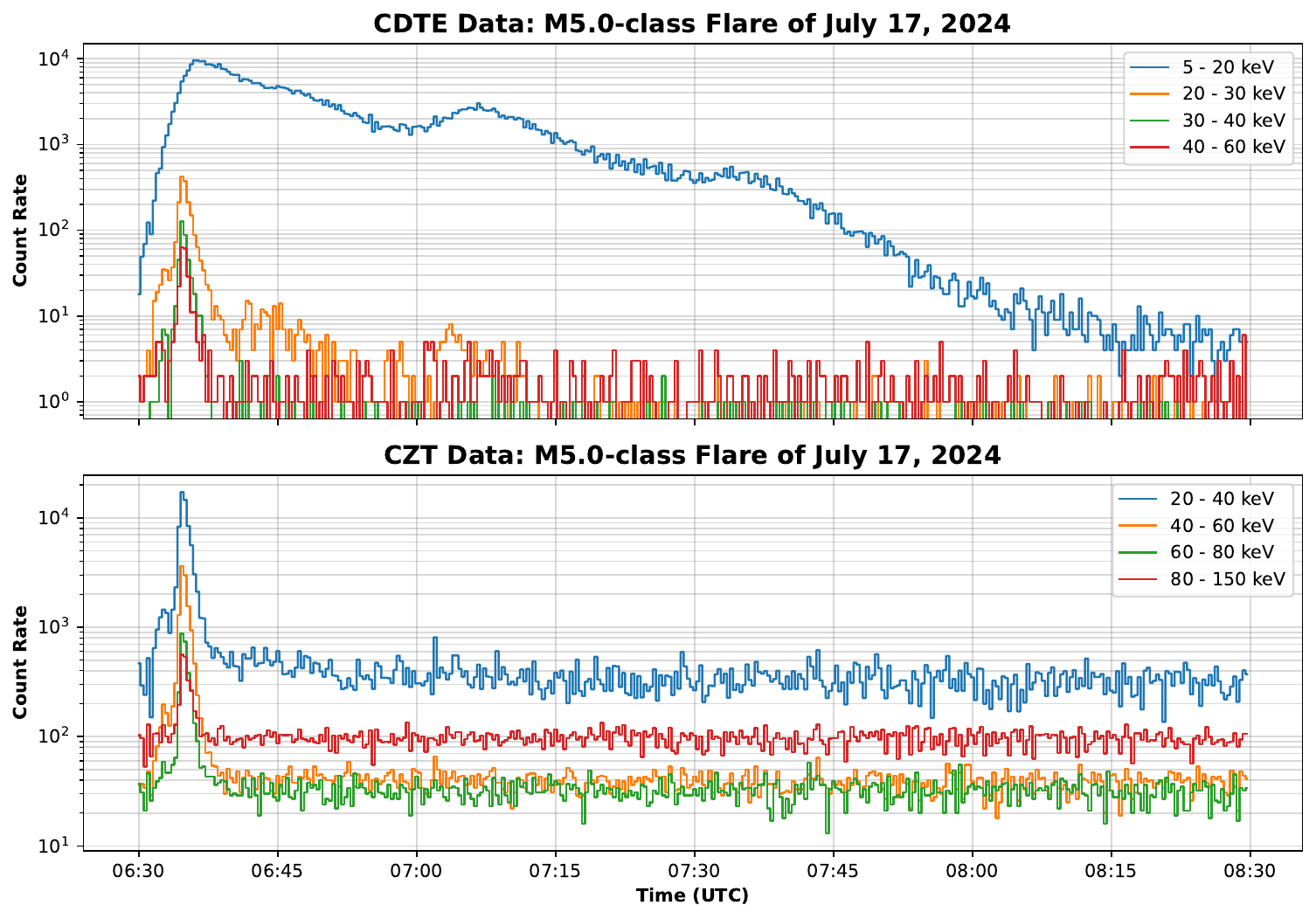}
    \captionof{figure}{The light curve product for the GOES M$5.0$
    class flare that occured on July~$17$, $2024$ -- plotted in
    different energy bands for the combined CdTe data on the top
    panel and for the combined CZT data on the bottom panel. The
    time period over which the spectrum in Figure~\ref{fig:specM}
    was integrated is $06$:$34$:$03$ -- $06$:$36$:$24$ UTC.}
    \label{fig:lcprod}    
\end{figure}

\begin{figure}[!ht]
    \centering
    \includegraphics[width=\columnwidth, viewport=69 18 832 432,clip]{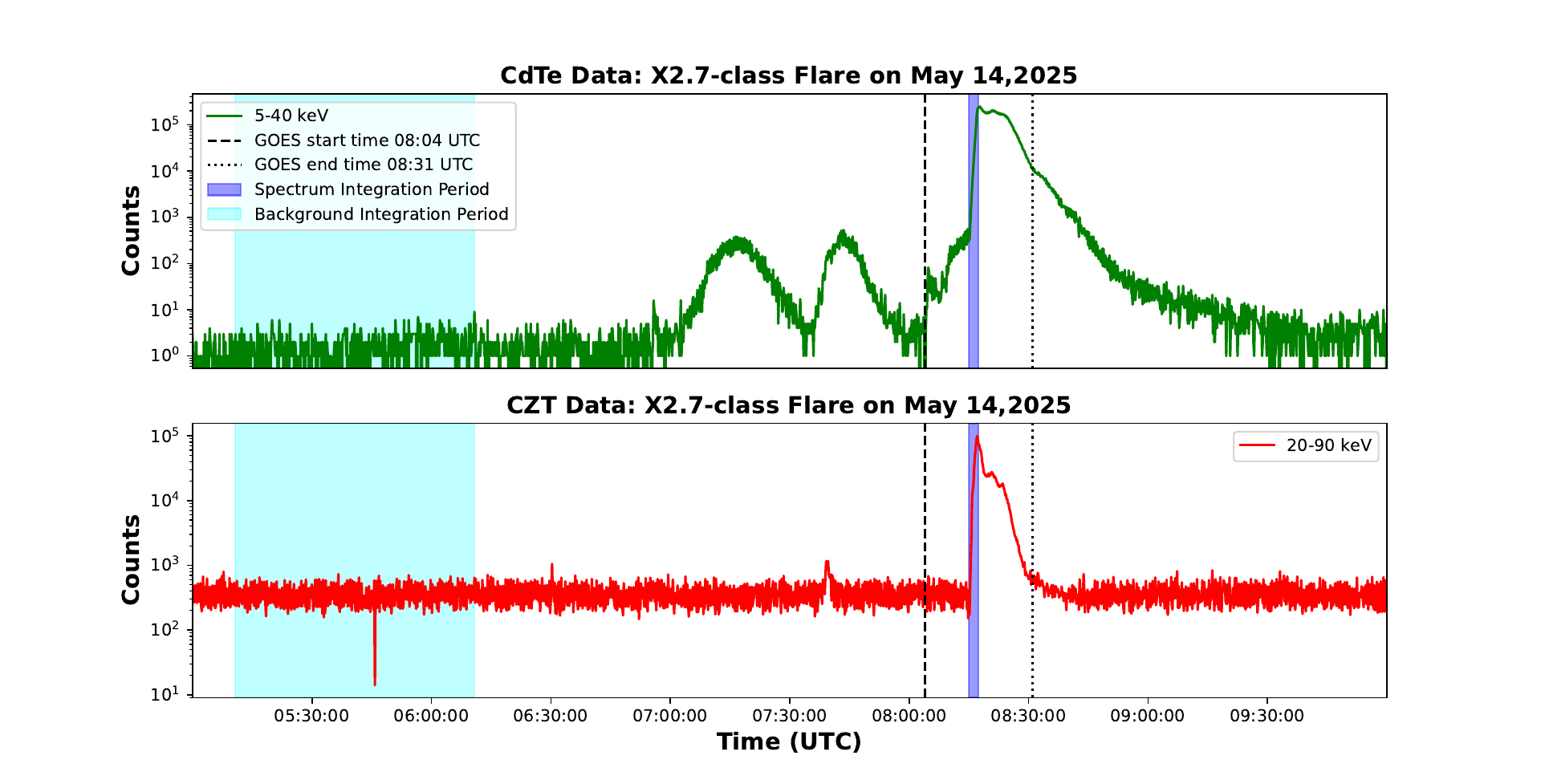}
    \captionof{figure}{Lightcurve of the GOES X$2.7$ class flare of
    May $14$, $2025$. The energy band of the CdTe and CZT data
    considered and the colours of the plot are same as that in
    Figure~\ref{fig:lcC}. The spectrum (plotted in
    Figure~\ref{fig:specX}) integration period (in purple band) is
    $08$:$15$:$01$ $-$ $08$:$17$:$22$ UTC, and the background
    integration time period (in cyan band) is $05$:$10$:$44$ $-$
    $06$:$10$:$48$ UTC.} 
    \label{fig:lcX}
\end{figure}

The lightcurves plotted in Figures~\ref{fig:lcC}, \ref{fig:lcprod}
and~\ref{fig:lcX} correspond to GOES C$6.5$-class, GOES M$5.0$-class
and GOES X$2.7$-class flares as mentioned earlier.
In the lightcurves  for the C-class flare and X-class flare, the GOES
start and end times are marked using vertical dashed and dotted lines.
The combined CdTe data is plotted for $5-40$~keV and combined CZT data
for $20-90$~keV energy ranges. Also marked in those figures are the
intervals of the data used for source spectrum and background spectrum
during the spectral analysis (see section~\ref{sec:spec}) in purple
and cyan bars respectively.

As mentioned in section~\ref{sec:proddetails} the lightcurve products
are generated for different sub-energy-bands at $1$~s cadence, and
are all saved as different bintable extensions of the Level-1
lightcurve FITS file for each of the detectors and these are for the
entire data-dump duration, which is typically $\sim$~$12$~hr. The
interactive lightcurve python utility disseminated in the `Other
Downloads' section on HEL1OS on PRADAN can be used with these
lightcurve FITS files to rebin the data, and also for selected time
and energy ranges -- thereby allowing one to restrict it for a flare
duration, or any other duration of interest. 
Illustrated in Figure~\ref{fig:lcprod} is one such plot created for
an M5.0 class flare that occured on July~$17$, $2024$ between
$\sim$~$6$:$30$~UTC and $\sim$~$8$:$30$~UTC. The plot is also
created after combining the data for two CdTe detectors and for
the two CZT detectors separately.

\subsubsection{Spectral Product Analysis}\label{sec:spec}
\begin{figure}[!ht]
    \centering
    \includegraphics[width=\columnwidth, trim=85 5mm 62mm 13mm,clip]{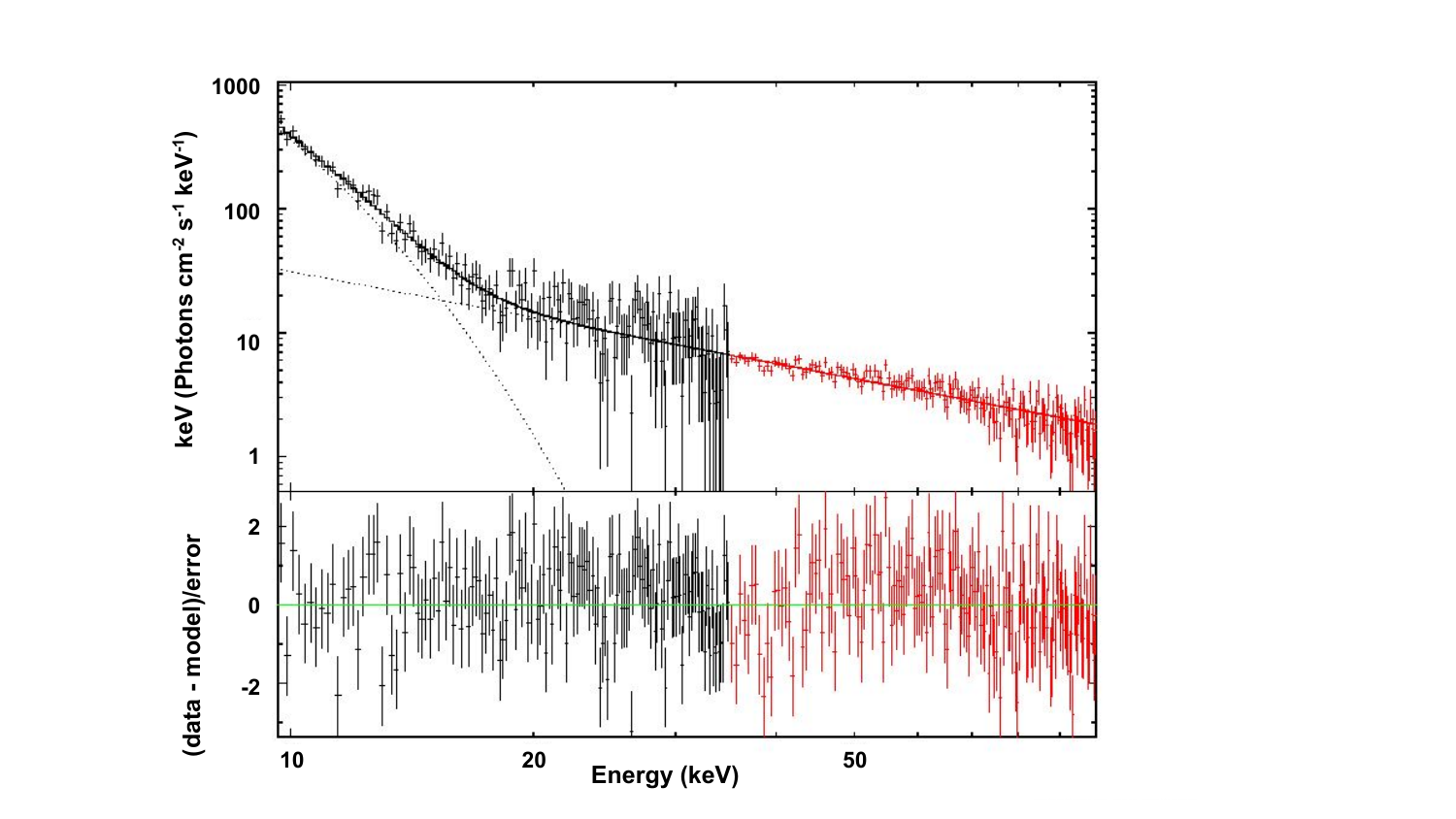}
    \caption{CdTe (black) and CZT (red) combined spectral fit on
    XSPEC for the flare corresponding to GOES C$6.5$ class on May
    $31$, $2025$ (for the source and background intervals marked
    in purple and cyan bars respectively in Figure~\ref{fig:lcC}).
    The reduced $\chi^2$ is $1.04$ and the photon power-law index is
    $2.23 \pm 0.03$.
    }
    \label{fig:specC}
\end{figure}

\begin{figure}[!ht]
    \centering
    \includegraphics[width=\columnwidth, trim=92 5mm 39mm 5mm,clip]{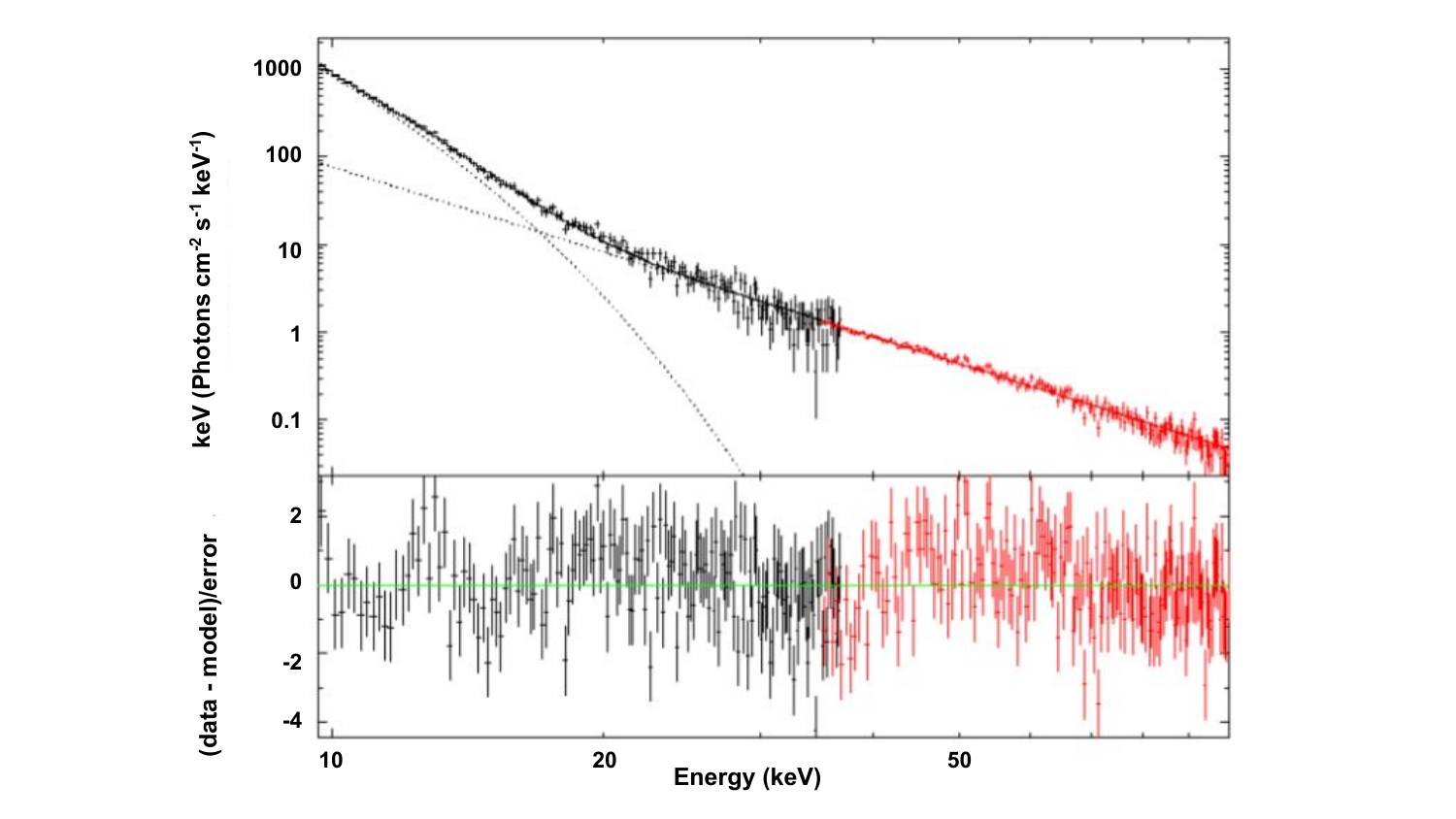}
    \caption{CdTe (black) and CZT (red) combined spectral fit for the
    flare corresponding to GOES M$5.0$ class on July~$17$, $2024$ (for
    the time period indicated in the caption of
    Figure~\ref{fig:lcprod}). The reduced $\chi^2$ is $1.58$ and the
    photon power-law index is $3.21 \pm 0.04$.
    }
    \label{fig:specM}
\end{figure}

\begin{figure}[!ht]
    \centering
    \includegraphics[width=\columnwidth, trim=80 5mm 62mm 13mm,clip]{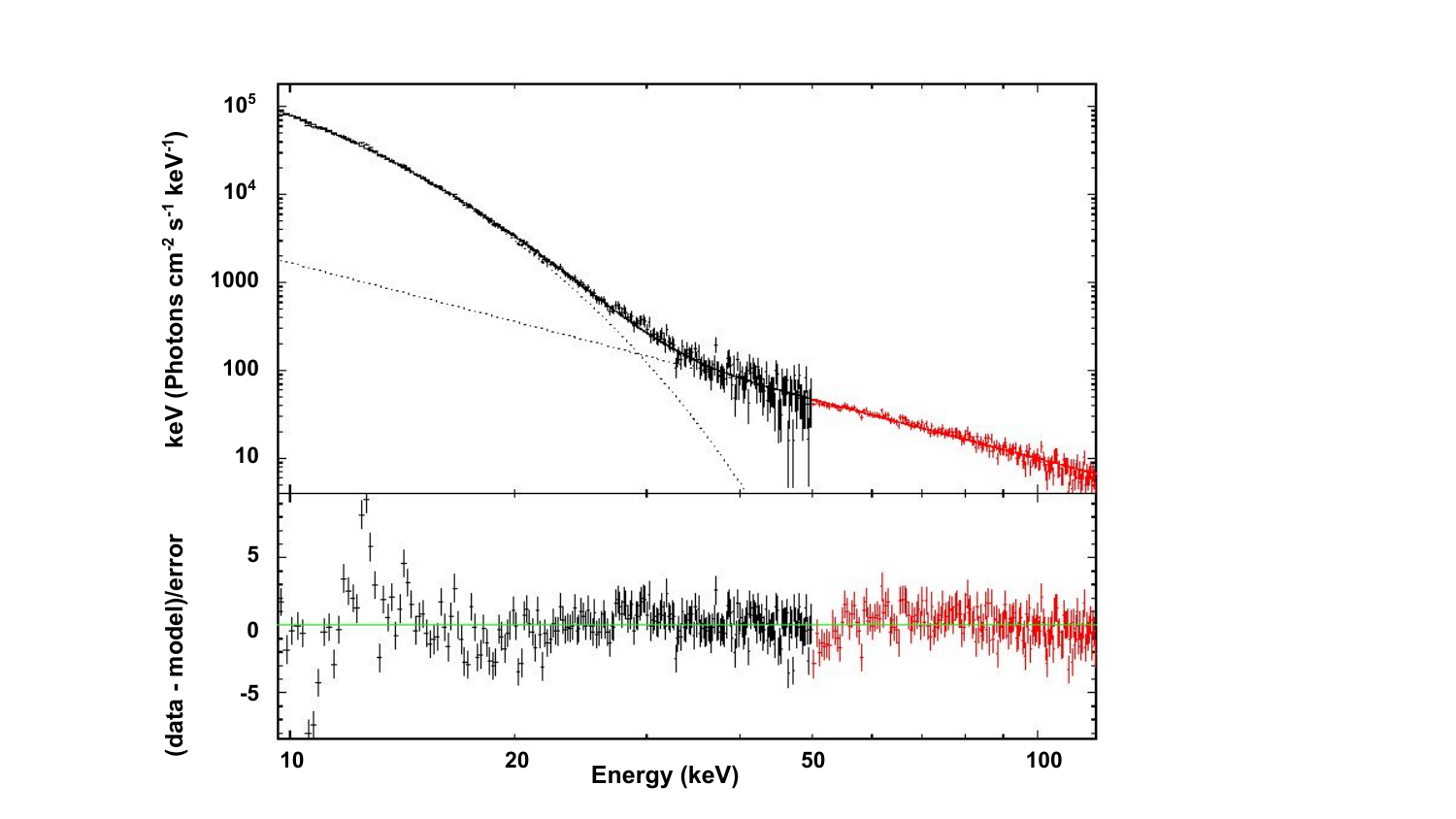}
    \caption{CdTe (black) and CZT (red) combined spectral fit for the
    flare corresponding to GOES X$2.7$ class on May $14$, $2025$ (for
    the source and background intervals marked in purple and cyan bars
    respectively in Figure~\ref{fig:lcX}). The reduced $\chi^2$ is
    $2.35$ and the photon power-law index is $3.22 \pm 0.03$.
    }
    \label{fig:specX}
\end{figure}

It is possible to undertake the spectral analysis on the HEL1OS data
with OSPEX, X-ray Spectral Fitting Package (XSPEC) or Sherpa tools.
Detailed information on the analysis using OSPEX has been provided
in the user manual which is made available on the PRADAN portal.
XSPEC allows for combined spectral fitting using data from multiple
instruments or detectors. The specifics of performing spectral
fitting with data products generated by the event list analysis
tools will be detailed in \cite{Manju:inpreparation} and will be
incorporated into the HEL1OS user manual later.
Here we show some results of the combined fitting done with the
CdTe as well as CZT detector data.
Figures~\ref{fig:specC}, \ref{fig:specM} and \ref{fig:specX} show
combined CdTe (in black) and CZT (in red) spectral
fits for representative C-class, M-class, and X-class solar flares
mentioned above respectively. The top panel of these
figures show the data points, model components and the post-fit model,
and their bottom panels show the residual of the respective fits.
These data points are picked during the impulsive phases of the
respective flares, mentioned in the captions of
Figures~\ref{fig:lcC}, \ref{fig:lcprod} and \ref{fig:lcX}, and also
highlighted in Figures~\ref{fig:lcC}, and \ref{fig:lcX}. 
The fits were performed using XSPEC models
\verb:apec+powerlaw: along with a \verb:constant: factor to handle
the cross-calibration difference between the two detectors. 
The constant for the reference detector (CdTe) was
fixed to unity, while the constant for the secondary detector (CZT)
was allowed to float freely during the fit. The cross-detector
normalisation constant is an empirical correction factor that absorbs
systematic discrepancies between the two independent detector systems
during joint forward-folded spectral modeling. Rather than being a
static instrument-to-instrument ratio, its value is driven by the
mathematical interplay between the chosen physical model, the
instrument response matrices, and the real-time photon flux. During
our combined detector solar flare fitting survey, we found that this
constant is uniquely dynamic due to the combination of typically
narrow energy overlap windows between the two detectors (3 to 5 keV)
and highly steep, fast-evolving spectra. A systematic error between
$2\%$ and $3\%$ has been included for the overall model fit for
both detectors. We have included the values of goodness-of-fit
statistic (reduced $\chi^2$) and the photon power-law indices for
each flare in the respective figure captions. The feature observed
in the residuals of all these flare spectra at $\sim$~$10-11$~keV is
likely instrumental in nature and can be accounted for by the user by
including a Gaussian line feature at that energy during the fit.

\section{Conclusion}

The HEL1OS payload on the Aditya-L1 mission is one of the few
experiments which provides an opportunity to observe Sun continuously
over a very wide hard X-ray energy band of $\sim$~$8$~keV $-$
$150$~keV with a stable background. The payload operational
parameters are fine-tuned and finalised. All the health
parameters are within operating limits, and the payload has been
functioning nominally providing spectro-temporal X-ray data of the
Sun. The rawdata of the payload are processed soon after they are
down-linked from the spacecraft and the processed data products are
made available to the solar community immediately. The science ready
data products consist of the ligthcurves and spectra. Some software
tools are also made available for specific scientific analysis
purposes. This paper illustrates the scientific potential and
capabilities of the HEL1OS experiment in temporal and spectral
analysis of solar flare activities.

\onecolumn

\begin{landscape}
\small 
\captionsetup{width=\linewidth}
\begin{xltabular}{\linewidth}{c c C{2cm} FFF FFr r}
    \caption{Flares with enhanced hard X-ray emission
    ($40$~keV~$-$~$150$~keV) detected in combined CZT detector from
    2024-07-17 up to 2025-08-05.} \label{tab:flarelist} \\
\toprule
\multirow{3}{*}{\textbf{No.}} & \multirow{3}{*}{\textbf{Date}} & \multirow{3}{*}{\textbf{Flare Class$^a$}} & \multicolumn{3}{c}{\textbf{GOES XRS (1--8 \AA)}$^b$} & \multicolumn{3}{c}{\textbf{CZT/HEL1OS (40--150 keV)}$^c$} & \textbf{CZT Fluence$^e$} \\
\cmidrule(lr){4-6} \cmidrule(lr){7-9}
 & & & \textbf{Start Time} & \textbf{Peak Time} & \textbf{End Time} & \textbf{Start Time} & \textbf{End Time} & \textbf{$\Delta T$$^d$} & \textbf{($\text{counts cm}^{-2}$)} \\
 & & & \textbf{(UTC)} & \textbf{(UTC)} & \textbf{(UTC)} & \textbf{(UTC)} & \textbf{(UTC)} & \textbf{(sec)} & \\
\midrule
\endfirsthead

\multicolumn{10}{c}{\tablename\ \thetable{} -- \textit{Continued from previous page}} \\
\toprule
\multirow{2}{*}{\textbf{No.}} & \multirow{2}{*}{\textbf{Date}} & \textbf{Flare Class$^a$} & \multicolumn{3}{c}{\textbf{GOES XRS (1--8 \AA)}$^b$} & \multicolumn{3}{c}{\textbf{CZT/HEL1OS (40--150 keV)}$^c$} & \textbf{CZT Fluence$^e$} \\
\cmidrule(lr){4-6} \cmidrule(lr){7-9}
 & &  & \textbf{Start Time} & \textbf{Peak Time} & \textbf{End Time} & \textbf{Start Time} & \textbf{End Time} & \textbf{$\Delta T$$^d$} & \textbf{($\text{counts cm}^{-2}$)} \\
 & & & \textbf{(UTC)} & \textbf{(UTC)} & \textbf{(UTC)} & \textbf{(UTC)} & \textbf{(UTC)} & \textbf{(sec)} & \\
\midrule
\endhead

\midrule \multicolumn{10}{r}{\textit{Continued on next page}} \\
\endfoot

\bottomrule
\multicolumn{10}{@{}l@{}}{%
  \parbox{\linewidth}{\vspace{4pt}\footnotesize
    $^a$ \textbf{Flare Class}: GOES flare classification in $1\text{--}8~\text{\AA}$ band.\par
    $^b$ \textbf{GOES XRS}: GOES timestamps from $1\text{--}8~\text{\AA}$ band.\par
    $^c$ \textbf{CZT/HEL1OS}: HEL1OS Hard X-ray start and end timestamps recorded in the $40\text{--}150\text{ keV}$ band by the combined $32\text{ cm}^2$ CZT detector pair.\par
    $^d$ \textbf{$\Delta T$}: Duration of the Hard X-ray emission in seconds ($T_{\text{end}} - T_{\text{start}}$).\par
    $^e$ \textbf{CZT Fluence}: Total counts $40\text{--}150\text{ keV}$  normalised combined CZT area, integrated across $\Delta T$ with units $\text{counts cm}^{-2}$.
  }%
} \\
\endlastfoot

1 & 2024-07-14 & X1.2 & 02:23:00 & 02:34:00 & 02:48:00 & 02:29:35 & 02:35:15 & 340 & 2186.75 \\
2 & 2024-07-16 & X1.9 & 13:11:00 & 13:26:00 & 13:36:00 & 13:15:35 & 13:27:55 & 740 & 8883.00 \\
3 & 2024-07-17 & M5.0 & 06:26:00 & 06:39:00 & 07:01:00 & 06:27:15 & 06:37:15 & 600 & 839.69 \\
4 & 2024-07-22 & M3.9 & 03:55:00 & 04:04:00 & 04:08:00 & 04:03:25 & 04:04:05 & 40 & 212.72 \\
5 & 2024-07-22 & C6.2 & 06:48:00 & 06:56:00 & 07:04:00 & 06:54:55 & 06:55:35 & 40 & 70.19 \\
6 & 2024-07-24 & M3.6 & 07:28:00 & 07:42:00 & 07:50:00 & 07:33:55 & 07:35:25 & 90 & 42.91 \\
7 & 2024-07-24 & M2.9 & 17:07:00 & 17:21:00 & 17:27:00 & 17:14:45 & 17:19:05 & 260 & 494.88 \\
8 & 2024-07-28 & M7.8 & 01:39:00 & 01:51:00 & 02:15:00 & 01:46:45 & 01:56:55 & 610 & 2588.38 \\
9 & 2024-07-29 & X1.5 & 02:00:00 & 02:37:00 & 02:43:00 & 02:32:25 & 02:37:35 & 310 & 1807.47 \\
10 & 2024-07-29 & M8.7 & 12:47:00 & 12:55:00 & 13:04:00 & 12:50:35 & 12:55:25 & 290 & 3179.44 \\
11 & 2024-07-31 & M5.4 & 18:07:00 & 18:37:00 & 18:57:00 & 18:20:15 & 18:35:55 & 940 & 3577.00 \\
12 & 2024-08-01 & M6.3 & 01:47:00 & 01:50:00 & 01:54:00 & 01:47:05 & 01:48:55 & 110 & 2412.66 \\
13 & 2024-08-01 & M4.0 & 04:35:00 & 04:40:00 & 04:45:00 & 04:38:05 & 04:42:25 & 260 & 1487.78 \\
14 & 2024-08-01 & M8.2 & 06:23:00 & 07:11:00 & 07:39:00 & 06:59:35 & 07:13:05 & 810 & 1182.59 \\
15 & 2024-08-02 & M2.1 & 07:51:00 & 07:57:00 & 08:03:00 & 07:57:35 & 07:57:55 & 20 & 110.34 \\
16 & 2024-08-03 & M7.3 & 18:29:00 & 18:38:00 & 18:44:00 & 18:36:55 & 18:39:35 & 160 & 1528.59 \\
17 & 2024-08-04 & M1.2 & 09:35:00 & 09:43:00 & 09:48:00 & 09:41:55 & 09:44:35 & 160 & 430.84 \\
18 & 2024-08-04 & M2.2 & 15:11:00 & 15:15:00 & 15:19:00 & 15:13:15 & 15:14:05 & 50 & 192.69 \\
19 & 2024-08-04 & M1.0 & 22:04:00 & 22:09:00 & 22:15:00 & 22:08:35 & 22:09:25 & 50 & 175.22 \\
20 & 2024-08-05 & M6.1 & 05:13:00 & 05:22:00 & 05:27:00 & 05:20:05 & 05:25:05 & 300 & 2476.66 \\
21 & 2024-08-05 & X1.1 & 15:18:00 & 15:27:00 & 15:32:00 & 15:22:35 & 15:28:25 & 350 & 1281.16 \\
22 & 2024-08-13 & M1.0 & 17:25:00 & 17:28:00 & 17:34:00 & 17:27:25 & 17:28:35 & 70 & 367.77 \\
23 & 2024-08-19 & M1.3 & 09:28:00 & 09:35:00 & 09:40:00 & 09:35:15 & 09:35:45 & 30 & 142.34 \\
24 & 2024-08-23 & M3.4 & 03:31:00 & 03:39:00 & 03:45:00 & 03:37:45 & 03:39:15 & 90 & 845.03 \\
25 & 2024-08-23 & C5.2 & 05:35:00 & 05:41:00 & 05:48:00 & 05:37:55 & 05:44:05 & 370 & 154.31 \\
26 & 2024-08-30 & M3.8 & 12:13:00 & 12:19:00 & 12:25:00 & 12:18:05 & 12:20:05 & 120 & 833.00 \\
27 & 2024-09-04 & M1.4 & 02:43:00 & 02:51:00 & 03:11:00 & 02:50:15 & 02:52:05 & 110 & 985.22 \\
28 & 2024-09-12 & X1.3 & 09:31:00 & 09:43:00 & 09:51:00 & 09:36:55 & 09:44:05 & 430 & 1976.73 \\
29 & 2024-09-12 & M6.8 & 14:31:00 & 14:41:00 & 14:47:00 & 14:40:45 & 14:42:05 & 80 & 70.31 \\
30 & 2024-09-14 & X4.5 & 15:13:00 & 15:29:00 & 18:47:00 & 15:16:35 & 15:47:45 & 1870 & 45162.69 \\
31 & 2024-10-01 & X7.1 & 21:58:00 & 22:20:00 & 23:59:00 & 22:08:35 & 22:20:05 & 690 & 12672.62 \\
32 & 2024-10-03 & C8.6 & 06:43:00 & 06:47:00 & 06:52:00 & 06:45:35 & 06:47:45 & 130 & 104.05 \\
33 & 2024-10-03 & X9.0 & 12:08:00 & 12:18:00 & 16:00:00 & 12:13:25 & 12:26:15 & 770 & 33002.97 \\
34 & 2024-10-07 & X2.1 & 19:02:00 & 19:13:00 & 19:31:00 & 19:04:05 & 19:17:15 & 790 & 8792.56 \\
35 & 2024-10-09 & X1.8 & 01:25:00 & 01:56:00 & 02:43:00 & 01:38:05 & 02:04:55 & 1610 & 17373.12 \\
36 & 2024-10-14 & M3.4 & 00:04:00 & 00:15:00 & 00:31:00 & 00:09:45 & 00:24:55 & 910 & 1878.72 \\
37 & 2024-10-15 & M2.1 & 18:00:00 & 18:21:00 & 18:40:00 & 18:07:25 & 18:32:25 & 1500 & 1310.69 \\
38 & 2024-10-17 & M2.4 & 04:53:00 & 05:02:00 & 05:13:00 & 05:00:05 & 05:03:05 & 180 & 557.59 \\
39 & 2024-10-18 & M4.7 & 23:13:00 & 23:25:00 & 23:38:00 & 23:16:45 & 23:27:55 & 670 & 644.81 \\
40 & 2024-10-24 & X3.3 & 03:30:00 & 03:57:00 & 04:28:00 & 03:38:25 & 04:09:05 & 1840 & 27309.72 \\
41 & 2024-10-25 & M1.1 & 07:23:00 & 07:31:00 & 07:38:00 & 07:30:25 & 07:31:25 & 60 & 659.44 \\
42 & 2024-10-31 & X2.0 & 21:12:00 & 21:20:00 & 22:40:00 & 21:14:45 & 21:22:05 & 440 & 16853.50 \\
43 & 2024-11-06 & X2.3 & 13:24:00 & 13:40:00 & 13:50:00 & 13:31:25 & 13:41:05 & 580 & 1067.59 \\
44 & 2024-12-12 & M2.2 & 17:31:00 & 17:40:00 & 17:50:00 & 17:39:35 & 17:40:55 & 80 & 421.97 \\
45 & 2024-12-13 & M1.0 & 14:15:00 & 14:23:00 & 14:31:00 & 14:23:05 & 14:30:15 & 430 & 168.66 \\
46 & 2024-12-20 & M2.5 & 11:15:00 & 11:18:00 & 11:22:00 & 11:17:15 & 11:18:15 & 60 & 240.19 \\
47 & 2024-12-21 & M1.9 & 00:33:00 & 00:37:00 & 00:42:00 & 00:36:55 & 00:38:05 & 70 & 492.16 \\
48 & 2024-12-22 & M1.0 & 04:03:00 & 04:09:00 & 04:14:00 & 04:09:05 & 04:09:45 & 40 & 85.16 \\
49 & 2024-12-23 & M8.9 & 11:06:00 & 11:10:00 & 11:16:00 & 11:08:05 & 11:11:35 & 210 & 1054.50 \\
50 & 2024-12-24 & M4.1 & 08:28:00 & 08:41:00 & 08:45:00 & 08:38:15 & 08:43:25 & 310 & 559.34 \\
51 & 2024-12-25 & M4.9 & 04:46:00 & 04:48:00 & 04:53:00 & 04:47:15 & 04:48:05 & 50 & 492.91 \\
52 & 2024-12-28 & M4.5 & 11:12:00 & 11:19:00 & 11:31:00 & 11:17:45 & 11:20:05 & 140 & 788.09 \\
53 & 2024-12-29 & M2.0 & 04:18:00 & 04:24:00 & 04:45:00 & 04:19:35 & 04:19:45 & 10 & 106.28 \\
54 & 2024-12-29 & M7.1 & 15:02:00 & 15:07:00 & 15:22:00 & 15:05:55 & 15:08:15 & 140 & 704.62 \\
55 & 2024-12-30 & X1.5 & 04:01:00 & 04:10:00 & 04:28:00 & 04:06:55 & 04:12:35 & 340 & 2705.19 \\
56 & 2024-12-30 & M5.0 & 16:45:00 & 16:54:00 & 17:01:00 & 16:53:05 & 16:54:45 & 100 & 249.06 \\
57 & 2025-01-03 & X1.2 & 11:29:00 & 11:39:00 & 11:49:00 & 11:32:55 & 11:38:35 & 340 & 355.56 \\
58 & 2025-01-03 & X1.1 & 22:32:00 & 22:41:00 & 22:51:00 & 22:35:45 & 22:43:05 & 440 & 1220.94 \\
59 & 2025-01-06 & M4.8 & 16:12:00 & 16:24:00 & 16:30:00 & 16:20:55 & 16:22:45 & 110 & 683.09 \\
60 & 2025-01-17 & M7.4 & 13:24:00 & 13:35:00 & 13:39:00 & 13:31:55 & 13:33:45 & 110 & 70.72 \\
61 & 2025-01-19 & M2.4 & 03:28:00 & 03:32:00 & 03:36:00 & 03:31:25 & 03:32:35 & 70 & 448.97 \\
62 & 2025-02-01 & M2.4 & 13:14:00 & 13:18:00 & 13:23:00 & 13:15:25 & 13:17:35 & 130 & 628.28 \\
63 & 2025-02-02 & M5.1 & 13:58:00 & 14:03:00 & 14:08:00 & 14:02:55 & 14:04:05 & 70 & 880.28 \\
64 & 2025-02-03 & M3.1 & 05:37:00 & 05:46:00 & 05:54:00 & 05:41:45 & 05:50:05 & 500 & 538.16 \\
65 & 2025-02-03 & M4.3 & 18:25:00 & 18:33:00 & 18:45:00 & 18:32:35 & 18:33:35 & 60 & 352.53 \\
66 & 2025-02-05 & C7.9 & 01:01:00 & 01:06:00 & 01:14:00 & 01:06:05 & 01:07:15 & 70 & 171.72 \\
67 & 2025-02-05 & M2.7 & 07:44:00 & 07:49:00 & 07:57:00 & 07:47:25 & 07:49:45 & 140 & 694.16 \\
68 & 2025-02-05 & C4.4 & 18:25:00 & 18:28:00 & 18:32:00 & 18:25:05 & 18:29:55 & 290 & 294.73 \\
69 & 2025-02-06 & M7.6 & 10:47:00 & 10:58:00 & 11:16:00 & 10:55:45 & 11:00:25 & 280 & 584.94 \\
70 & 2025-02-23 & M4.9 & 02:00:00 & 02:11:00 & 02:23:00 & 02:07:55 & 02:13:35 & 340 & 958.16 \\
71 & 2025-02-23 & X2.0 & 19:22:00 & 19:28:00 & 19:34:00 & 19:23:25 & 19:29:35 & 370 & 2945.66 \\
72 & 2025-02-24 & M3.3 & 06:53:00 & 07:01:00 & 07:08:00 & 07:00:05 & 07:01:45 & 100 & 696.97 \\
73 & 2025-03-11 & M1.1 & 13:03:00 & 13:05:00 & 13:06:00 & 13:03:25 & 13:04:05 & 40 & 275.09 \\
74 & 2025-04-29 & M1.7 & 05:10:00 & 05:13:00 & 05:17:00 & 05:10:25 & 05:16:55 & 390 & 986.34 \\
75 & 2025-04-30 & M2.0 & 07:41:00 & 07:49:00 & 07:56:00 & 07:48:15 & 07:50:15 & 120 & 386.91 \\
76 & 2025-05-13 & X1.2 & 15:25:00 & 15:35:00 & 15:44:00 & 15:33:15 & 15:37:05 & 230 & 4627.12 \\
77 & 2025-05-14 & X2.7 & 08:04:00 & 08:18:00 & 08:31:00 & 08:15:45 & 08:24:45 & 540 & 5280.34 \\
78 & 2025-05-14 & M7.7 & 11:04:00 & 11:14:00 & 11:32:00 & 11:13:05 & 11:15:45 & 160 & 270.75 \\
79 & 2025-05-14 & M4.7 & 17:59:00 & 18:06:00 & 18:18:00 & 18:04:35 & 18:07:45 & 190 & 920.09 \\
80 & 2025-05-25 & X1.1 & 01:40:00 & 01:50:00 & 01:57:00 & 01:47:25 & 01:53:05 & 340 & 1391.31 \\
81 & 2025-05-25 & M8.9 & 16:18:00 & 16:28:00 & 16:36:00 & 16:26:15 & 16:30:45 & 270 & 1022.17 \\
82 & 2025-05-31 & M4.5 & 08:08:00 & 08:17:00 & 08:23:00 & 08:16:15 & 08:19:15 & 180 & 1135.78 \\
83 & 2025-05-31 & C6.5 & 13:21:00 & 13:27:00 & 13:30:00 & 13:26:55 & 13:27:45 & 50 & 475.72 \\
84 & 2025-06-15 & M8.4 & 17:45:00 & 18:07:00 & 18:25:00 & 17:49:25 & 18:24:55 & 2130 & 10897.75 \\
85 & 2025-06-16 & M6.3 & 09:17:00 & 09:35:00 & 09:48:00 & 09:33:25 & 09:36:35 & 190 & 922.31 \\
86 & 2025-06-19 & X1.9 & 23:37:00 & 23:46:00 & 23:54:00 & 23:45:55 & 23:47:05 & 70 & 269.16 \\
87 & 2025-08-05 & M4.4 & 15:46:00 & 15:51:00 & 15:58:00 & 15:50:15 & 15:52:45 & 150 & 491.28 \\

\end{xltabular}
\end{landscape}

\twocolumn

\section*{Acknowledgement}
The authors are grateful to the valuable inputs from the two anonymous
reviewers -- that helped in improving the quality of this paper.
The authors are thankful to the Group Head, Space Astronomy Group,
the Deputy Director, Payload Data Management and Space Astronomy area,
and the Director at the U. R. Rao Satellite Centre for the support to
carry out this work.  The Indian Space Research Organisation (ISRO)
funded, managed and facilitated the overall project.

\section*{Data Availability}
The HEL1OS data products are disseminated on the PRADAN web portal of
ISSDC at URL: https://pradan1.issdc.gov.in/al1/.

\bibliography{HEL1OS_pipeline_processing_rev1}

\end{document}